\documentclass[english,11pt]{article}
\usepackage[utf8]{inputenc}
\usepackage{graphicx}
\usepackage{color}
\usepackage{float}
\usepackage{slashed}
\usepackage{amsthm,amsmath,amssymb,mathrsfs}
\usepackage{setspace}
\allowdisplaybreaks
\usepackage{jheppub}
\usepackage{xcolor} 
\usepackage{babel}
\usepackage{booktabs}
\def\thefootnote{\fnsymbol{footnote}}

\def\beq{\begin{equation}}
\def\eeq{\end{equation}}

\begin{document}
\title{Neutrino forces in neutrino backgrounds}
\author[a]{Mitrajyoti Ghosh,}
\emailAdd{mg2338@cornell.edu}
\author[a]{Yuval Grossman,}
\emailAdd{yg73@cornell.edu}
\author[b]{Walter Tangarife,}
\emailAdd{wtangarife@luc.edu}
\author[c,d]{Xun-Jie Xu,} 
\emailAdd{xuxj@ihep.ac.cn}
\author[c,d]{Bingrong Yu\footnote{corresponding author}
} 
\emailAdd{yubr@ihep.ac.cn}
\affiliation[a]{Department of Physics, LEPP, Cornell University, Ithaca, NY 14853, USA}
\affiliation[b]{Department of Physics, Loyola University Chicago, Chicago, IL 60660, USA} \affiliation[c]{Institute of High Energy Physics, Chinese Academy of Sciences, Beijing 100049, China}
\affiliation[d]{School of Physical Sciences, University of Chinese Academy of Sciences, Beijing 100049, China}

\abstract{
\looseness=-1 The Standard Model predicts a long-range force, proportional to $G_F^2/r^5$, between fermions due to the exchange of a pair of neutrinos.
This quantum force  is feeble and has not been observed yet.
In this paper, we compute this force in the presence of neutrino backgrounds, both for isotropic and directional background neutrinos.
We find that for the case of directional background the force can have a $1/r$ dependence and it can be significantly enhanced compared to the vacuum case.
In particular, background effects
caused by reactor,  solar, and supernova neutrinos
enhance the force by many orders of magnitude. The enhancement, however, occurs only in the direction parallel to the direction of the background neutrinos.
We discuss the experimental prospects of detecting the neutrino force in neutrino backgrounds and find that the effect is close
to the available sensitivity of the current fifth force experiments. 
Yet, the angular spread of the neutrino flux and that of the test masses reduce the strength of this force. 
The results are encouraging and a detailed experimental study is called for to  check if the effect can be probed. 
}

\maketitle

\def\thefootnote{\arabic{footnote}}
\section{Introduction}
\label{sec:intro}

\looseness=-1 
It is well known that classical forces, like the Coulomb potential, can be derived from a $t$-channel mediator-exchange diagram in quantum field theory. The same treatment can be applied to the exchange of massive gauge bosons and scalars, resulting in a Yukawa potential.
To obtain a classical force, the mediator of the force must be a boson. However, a pair of fermions behaves as an effective scalar and can mediate long-range forces. Such forces are sometimes called ``quantum forces.'' Quantum forces have been studied extensively in the literature, for example, see~\cite{Hsu:1992tg, Brax:2017xho,Fichet:2017bng,Costantino:2019ixl}, in an attempt to both test the Standard Model (SM) and to probe new physics beyond.

\begin{figure}
	\centering
\includegraphics{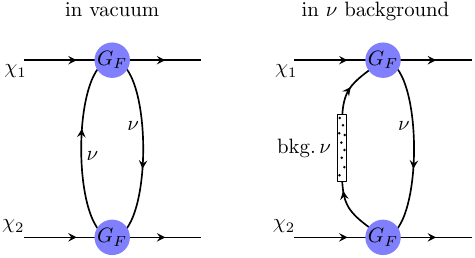}
\caption{\label{fig:feyn}A diagrammatic explanation of neutrino forces in the vacuum (left) and in a neutrino background (right). The background effect can be taken into account by replacing one of the neutrino propagators with a background-modified propagator (bkg.\,$\nu$), which can be computed in finite temperature field theory. The effect can be physically interpreted as absorbing a neutrino from the background and then returning it back to the background.
}
	
\end{figure}

In the SM, the force between fermions due to  neutrino pair exchange is also well studied. Since neutrinos are very light, the force mediated by them is long range, without any significant exponential suppression with distance. Neutrino forces are generated by the exchange of a neutrino-antineutrino
pair between two particles, as shown in the left panel of Fig.~\ref{fig:feyn}.
The original idea of the neutrino-mediated force can be traced back to Feynman, who tried to explain the $1/r$ gravity as an emergent phenomenon due to the exchange of two neutrinos when taking into account multi-body effects~\cite{Feynmangravitation}. 
%It is so interesting and intuitive that here we ``rediscover" the $1/r$ behavior of the force when considering the effects from a directional neutrino background.
Previous
calculations of such forces in vacuum were first carried out in Refs.~\cite{Feinberg:1968zz,Feinberg:1989ps,Hsu:1992tg} using the dispersion
technique for massless neutrinos. Later, the effects of neutrino masses~\cite{Grifols:1996fk} and flavor mixing~\cite{Lusignoli:2010gw,LeThien:2019lxh,Costantino:2020bei} were included, which in principle can be used to determine the nature of
neutrinos~\cite{Segarra:2020rah,Costantino:2020bei}, namely, whether neutrinos are Dirac or Majorana particles. The study of neutrino forces in the framework of effective field theories was carried out in Ref.~\cite{Bolton:2020xsm}.

Neutrino forces have important cosmological and astrophysical effects, such as the stability of neutron stars~\cite{Fischbach:1996qf,Smirnov:1996vj,Abada:1996nx,Kachelriess:1997cr,Kiers:1997ty,Abada:1998ti,Arafune:1998ft} and the impact on dark matter in the early universe~\cite{Orlofsky:2021mmy,Coy:2022cpt}. Recently, the calculation of neutrino forces went beyond the  four-fermion contact interaction and a general formula describing the short-range behavior of neutrino forces was derived~\cite{Xu:2021daf}.

While theoretically we know that the force should be there, it has never been confirmed experimentally. The reason is that the force is very weak. The fact that it is second order in the weak interaction makes it proportional to $G_F^2$. In the limit of massless neutrinos, it is explicitly
\begin{equation}
V(r) \sim \frac{G_F^2}{r^5}\;,
\end{equation}
where $G_F=1.166\times 10^{-5} {\rm GeV}^{-2}$ is the Fermi constant and $r$ is the distance between the two particles.
Thus, already at distances longer than about a nanometer, the neutrino force is smaller than the gravitational force between elementary particles.

Confirming the neutrino force experimentally would be interesting for several reasons. First, it would establish an exciting prediction of quantum field theory that remains untested. Second, it would enable us to probe the neutrino sector of the SM since the neutrino force is sensitive to the absolute masses of the neutrinos. Also, it provides a test of the electroweak interaction and may serve as a probe of new physics beyond the SM. Lastly, it would enable us to look for other quantum forces that may be present due to yet undiscovered light particles~\cite{Grifols:1994zz,Ferrer:1998ue,Fichet:2017bng,Brax:2017xho,Costantino:2019ixl,Banks:2020gpu}. 

Given that the neutrino force is so feeble, we need to look for novel ways to probe it. One such idea was put forward in~\cite{Ghosh:2019dmi}, which pointed out that the neutrino force provides the leading long-range parity-violation effect in the SM. Thus, it is natural to look for such effects. Yet even this seems too small to be probed experimentally.

In this paper, we explore a different path:  the neutrino force in the presence of an intense neutrino background, as shown in the right panel of Fig.~\ref{fig:feyn}. The presence of the background can significantly increase the strength of the interaction. 
In fact, the effect of a neutrino background was studied before, for the cosmic neutrino background (C$\nu$B), in Refs.~\cite{Horowitz:1993kw,Ferrer:1998ju,Ferrer:1999ad}. However, the effect in this case is small because the 
number density of the cosmic neutrinos is very small today.

In this work, we focus on scenarios where the background is much more dense; in particular, for solar and reactor neutrinos. On the theoretical level, this differs from the case of C$\nu$B in that the background is not spherically symmetric. This results in a preferred direction, providing a fundamentally different signal than that of the vacuum and C$\nu$B cases.

Numerically, we find that the effect of reactor and solar neutrinos is remarkably significant and can enhance the signal by more than 20 orders of magnitude. In particular, the encouraging result is that the effect is close to the available sensitivity of fifth-force experimental searches. Thus, we hope that using the effect of background neutrinos will enable us to probe the neutrino force.

The paper is organized as follows. In Sec.~\ref{sec:formalism}, we set up the general formalism to calculate the neutrino force in an arbitrary neutrino background. After applying this formalism to the case of C$\nu$B in Sec.~\ref{sec:CnuB}, we calculate the neutrino force in a directional neutrino flux background in Sec.~\ref{sec:reactorsolor}. In Sec.~\ref{sec:exp}, we discuss the detection of neutrino forces in neutrino backgrounds and compare our theoretical results with the experimental sensitivities. Our main conclusions are summarized in Sec.~\ref{sec:summary}. The technical details are expanded in the appendices.

\section{Formalism}
\label{sec:formalism}
In this section, we introduce the general formalism to compute
neutrino forces between two fermions in a general neutrino background. Consider a four-fermion interaction with two Dirac neutrinos (for the case of Majorana neutrinos, see Sec.~\ref{subsec:majorana case}) and two fermions:
\begin{eqnarray}
{\cal L}_{{\rm int}}=-\frac{G_{F}}{\sqrt{2}}\left[\bar{\nu}\gamma^{\mu}\left(1-\gamma_{5}\right)\nu\right] \left[\bar{\chi}\gamma_{\mu}\left(g_{V}^{\chi}+g_{A}^{\chi}\gamma_{5}\right)\chi\right]\;,\label{eq:lagDirac}
\end{eqnarray}
where $G_{F}$ is the Fermi constant, $\nu$ denotes a Dirac neutrino 
with mass $m_{\nu}$, $\chi$ is a generic fermion in or beyond the SM with mass $m_{\chi}$, $g_{V}^{\chi}$ and $g_{A}^{\chi}$ are effective vector and axial couplings of $\chi$ to the neutrinos, obtained from integrating out heavy weak bosons. 

Before we start, we note the following:
\begin{enumerate}
\item We work in the non-relativistic (NR)  limit, i.e, the velocity of the interacting fermions $v \ll 1$ . The description of particle scattering via a potential $V(\mathbf{r})$ is accurate only in this limit.
\item Throughout our work, we only consider the spin-independent part of the potential. The reason is that the spin-dependent parts are usually averaged out when neutrino forces are added at macroscopic scales. The spin-independent part of the potential only depends on the vector coupling $g_{V}^{\chi}$. In Table~\ref{table:coupling in SM}, we collect the values of $g_{V}^{\chi}$ in the SM~\cite{Giunti:2007ry}. When $\chi$ is the proton or the neutron, $g_{V}^{\chi}$ can be obtained by simply summing over the vector couplings to the quarks.
\end{enumerate}

In vacuum, the diagram in the left panel of Fig.~\ref{fig:feyn} leads to a long-range
force that we can describe by an effective potential proportional
to $r^{-5}$ in the massless-neutrino limit, $r$ being the distance of the two external particles. More explicitly, the spin-independent part of the neutrino potential between two fermions $\chi_1$ and $\chi_2$ in that limit reads
\begin{eqnarray}
V_{0}(r)=\frac{G_{F}^{2}g^1_{V} g^2_V}{4\pi^{3}}\frac{1}{r^{5}}\qquad(m_{\chi_{1,2}}^{-1}\ll r\ll m_{\nu}^{-1})\;.\label{eq:vacuumpotential}
\end{eqnarray}
Here, we use $g^1_V \equiv g^{\chi_1}_V$ and $g^2_V \equiv g^{\chi_2}_V$ to simplify the notation. Note that, for $r\gg1/m_{\nu}$, the potential is exponentially
suppressed by $e^{-2m_\nu r}$~\cite{Grifols:1996fk}, while the NR approximation of $\chi$ becomes invalid as $r$ approaches $m_{\chi_{1,2}}^{-1}$. The short-range behavior of neutrino forces was first investigated in Ref.~\cite{Xu:2021daf}.

%%%%%%%%%%%%%%%%%%%%%%%%%%%% Table 1 %%%%%%%%%%%%%%%%%%%%%%%%%%
\renewcommand\arraystretch{1.2}
\begin{table}[t!]
	\centering
	\begin{tabular}{c|c|c|c|c|c}
		\hline\hline
		neutrino flavor & $\chi=e$ & $\chi=u$ & $\chi=d$ &  $\chi=$proton & $\chi=$neutron\\
		\hline
		$\nu_e$ & $\frac{1}{2}+2 s_{W}^2$& $\frac{1}{2}-\frac{4}{3} s_{W}^2$ & $-\frac{1}{2}+\frac{2}{3} s_{W}^2$  &  $\frac{1}{2}-2s_W^2$   & $-\frac{1}{2}$\\
		\hline
		$\nu_\mu$, $\nu_\tau$ &  $-\frac{1}{2}+2 s_{W}^2$ &  $\frac{1}{2}-\frac{4}{3} s_{W}^2$ & $-\frac{1}{2}+\frac{2}{3} s_{W}^2$  &  $\frac{1}{2}-2s_W^2$   & $-\frac{1}{2}$\\
		\hline\hline
	\end{tabular}
	\caption{\label{table:coupling in SM}Values of the vector coupling $g_V$ in Eq.~(\ref{eq:lagDirac}) in the SM. 
		%		for  $\chi$ being electron, proton and neutron~\cite{Giunti:2007ry}. 
		Here $s_W\equiv \sin\theta_W$ is the sine of the Weinberg angle.}
\end{table}	
\renewcommand\arraystretch{1.0}
%%%%%%%%%%%%%%%%%%%%%%%%%%%%%%%%%%%%%%%%%%%%%%%%%%%%%

In a neutrino background with finite neutrino number density or temperature,
the neutrino propagator should be modified, as shown on the right panel of Fig.~\ref{fig:feyn}. 
The modified propagator is often derived in the real-time formalism in finite temperature field
 theory (for a detailed review, see Refs.~\cite{Landsman:1986uw,Notzold:1987ik,Quiros:1999jp,Kapusta:2006pm,Laine:2016hma}.
	Also, see Appendix~\ref{sec:background-effect} for a simple and
	pedagogical re-derivation of the modified propagator.)
We then have:
\begin{eqnarray}
\!\!\!\!\!\!S_{\nu}(k)  =  \left(\slashed{k}+m_{\nu}\right) \left\{ \frac{i}{k^{2}-m_{\nu}^{2}+i\epsilon}-2\pi\delta\left(k^{2}-m_{\nu}^{2}\right)\left[\Theta\left(k^{0}\right)n_{+}\left(\mathbf{k}\right)+\Theta\left(-k^{0}\right)n_{-}\left(\mathbf{k}\right)\right]\right\} ,\label{eq:S}
\end{eqnarray}
where $\epsilon\to0^{+}$, $\Theta$ is the Heaviside theta function,
and $n_{\pm}\left(\mathbf{k}\right)$ denote the momentum distributions
of the neutrinos and anti-neutrinos respectively, such that the integrals $\int n_{\pm}\left(\mathbf{k}\right)d^{3}\mathbf{k}/(2\pi)^{3}$
correspond to their respective number densities. The first part is
the usual fermion propagator in vacuum while the second part accounts
for the background effect. The second part might seem counter-intuitive in the sense that the
	Dirac delta function requires the neutrino to be on-shell while, in
	Fig.~\ref{fig:feyn}, this on-shell neutrino is used to connect two
	spatially separated particles. To understand this effect, one should
	keep in mind that when $k$ in Eq.~\eqref{eq:S} is fixed, the uncertainty principle dictates that the neutrino
	cannot be localized and is spread out over space. So theoretically, the propagator's second (background) term,  just like the
	vacuum part,
	can mediate momentum over a large distance. 

According to the Born approximation, the effective potential is the
Fourier transform of the low-energy elastic scattering amplitude of $\chi_1$ with $\chi_2$,
\begin{equation}
V(\mathbf{r})=-\int\frac{d^{3}\mathbf{q}}{\left(2\pi\right)^{3}}e^{i\mathbf{q}\cdot\mathbf{r}}{\cal A}(\mathbf{q})\thinspace.\label{eq:V_basic}
\end{equation}
Here, ${\cal A}(\mathbf{q})$ is the scattering amplitude in the NR limit, which should
be  computed by integrating the neutrino loop in Fig.~\ref{fig:feyn}
using the modified neutrino propagator in Eq.~\eqref{eq:S}: 
\begin{eqnarray}
i{\cal A}(q)=\frac{G_{F}^{2} g^1_V g^2_V}{2}\int\frac{d^{4}k}{\left(2\pi\right)^{4}}{\rm Tr}\left[\gamma^{0}\left(1-\gamma_{5}\right)S_{\nu}(k)\gamma^{0}\left(1-\gamma_{5}\right)S_{\nu}\left(k+q\right)\right].\label{eq:amplitude}
\end{eqnarray}
Using the NR approximation we have $q\approx(0,\mathbf{q})$, thus the amplitude ${\cal A}$ only depends on the three-momentum $\mathbf{q}$.
Substituting Eq.~\eqref{eq:S} into Eq.~\eqref{eq:amplitude}, one
can see that when both neutrino propagators in Eq.~(\ref{eq:amplitude})
take the first term in the curly bracket of Eq.~\eqref{eq:S}, it
leads to the vacuum potential $V_{0}(r)$. When both propagators take
the second term, the result vanishes, as we show in Appendix~\ref{sec:integral}.
The background effect comes from cross terms, being proportional to
$n_{\pm}$. We denote the background contribution to ${\cal A}(\mathbf{q})$
by ${\cal A}_{{\rm bkg}}(\mathbf{q})$ and, correspondingly, the contribution
to $V(\mathbf{r})$ by $V_{{\rm bkg}}(\mathbf{r})$:
\begin{equation}
{\cal A}(\mathbf{q})={\cal A}_{0}(\mathbf{q})+{\cal A}_{{\rm bkg}}(\mathbf{q})\thinspace,\qquad V(\mathbf{r})=V_{0}(\mathbf{r})+V_{{\rm bkg}}(\mathbf{r})\thinspace.\label{eq:x-12}
\end{equation}
%Notice that the interference term between the vacuum amplitude ${\cal A}_{0}(\mathbf{q})$ and the background amplitude ${\cal A}_{{\rm bkg}}(\mathbf{q})$ would contribute to the cross section but has no influence on the effective potential.
Notice that there is no interference between the vacuum and the background amplitudes in our calculation because, unlike computing cross sections, here we do not need to square the total amplitude.  
The background contribution ${\cal A}_{{\rm bkg}}(\mathbf{q})$, after
some calculations in Appendix~\ref{sec:integral}, reduces to 
\begin{eqnarray}
{\cal A}_{{\rm bkg}}(\mathbf{q})=4G_{F}^{2} g^1_V g^2_V\int\frac{d^{3}\mathbf{k}}{\left(2\pi\right)^{3}}\frac{n_{+}\left(\mathbf{k}\right)+n_{-}\left(\mathbf{k}\right)}{2E_{k}}\left[\frac{2\left|\mathbf{k}\right|^{2}+m_{\nu}^{2}+\mathbf{k}\cdot\mathbf{q}}{2\mathbf{k}\cdot\mathbf{q}+\left|\mathbf{q}\right|^{2}}+\left(\mathbf{k}\to-\mathbf{k}\right)\right].\label{eq:Abkg}
\end{eqnarray}

For isotropic distributions (e.g.~cosmic neutrino background, diffuse
supernova neutrino background), $n_{\pm}$ are independent of the
direction of the momentum, i.e., $n_{\pm}(\mathbf{k})=n_{\pm}(\kappa)$ with $\kappa\equiv\left|\mathbf{k}\right|$, leading to an isotropic ${\cal A}_{{\rm bkg}}$
and hence an isotropic $V_{{\rm bkg}}$. In this case, the angular
part of the above integral can be integrated out analytically, resulting
in the following expression for $V_{{\rm bkg}}$: 
\begin{equation}
V_{{\rm bkg}}(r)=-\frac{G_{F}^{2} g^1_V g^2_V}{4\pi^{3}r^{4}}\int_{0}^{\infty}d\kappa\,\kappa\,\frac{n_{+}\left(\kappa\right)+n_{-}\left(\kappa\right)}{\sqrt{\kappa^{2}+m_{\nu}^{2}}}\left[\left(1+m_{\nu}^{2}r^{2}\right)\sin\left(2\kappa r\right)-2\kappa r\cos\left(2\kappa r\right)\right].\label{eq:V_iso}
\end{equation}

Up to now, we have not used any specific neutrino distributions. In
what follows, we   apply the above formulae to specific forms of
$n_{\pm}$ and compute the corresponding potentials.

\section{Neutrino forces with isotropic  neutrino background}
\label{sec:CnuB}

We now discuss the case where the neutrino background is isotropic and focus on a thermal-like distribution. In particular, this applies to the cosmic neutrino background (C$\nu$B), which motivates this section. 

The existence of isotropic C$\nu$B today, with a temperature around 1.9~K and number density about $56/{\rm cm}^{3}$ per flavor, is one of the most solid predictions from big bang cosmology ~\cite{Weinberg2008cosmology}.
The temperature correction to neutrino forces in the C$\nu$B was first calculated in Ref.~\cite{Horowitz:1993kw} with the neutrino momentum distribution to be
\begin{eqnarray}
	\label{eq:MBdistribution1}
	n_{\pm}\left(\mathbf{k},T\right)={\rm exp}\left[\left(\pm\mu-\kappa\right)/T\right]\quad{\rm with}\quad 
	\kappa\equiv\left|\mathbf{k}\right|\;,
\end{eqnarray}
where $\mu$ and $T$ are the chemical potential and temperature of the C$\nu$B. 

Ref.~\cite{Horowitz:1993kw} studied the case of Dirac neutrinos in the massless ($m_\nu=0$) and NR ($m_\nu\gg T$) limit. Later, the background effects of the C$\nu$B on neutrino forces were further studied in Ref.~\cite{Ferrer:1998ju,Ferrer:1999ad}. In Ref.~\cite{Ferrer:1998ju} the neutrino distribution was taken to be a standard Boltzmann distribution, 
\begin{eqnarray}
	\label{eq:MBdistribution2}
	n_{\pm}\left(\mathbf{k},T\right)={\rm exp}\left[\left(\pm\mu-E_k\right)/T\right]\quad{\rm with}\quad
	E_k=\sqrt{\left|\mathbf{k}\right|^2+m_\nu^2}\;,
\end{eqnarray}
and the complete expressions of the background potential $V_{\rm bkg}(r)$ were given for both Dirac and Majorana neutrinos. The massless limit of the result in Ref.~\cite{Ferrer:1998ju} matches that in Ref.~\cite{Horowitz:1993kw}. However, the results of the massive case are very different. In particular, the expression of $V_{\rm bkg}(r)$ in Ref.~\cite{Ferrer:1998ju} is exponentially suppressed at large distances, $V_{\rm bkg}(r)\sim e^{-2m_\nu r}$ (for $r\gg 1/m_\nu$), while that in Ref.~\cite{Horowitz:1993kw} is not, $V_{\rm bkg}(r)\sim m_\nu/(Tr^5)$ (for $r\gg 1/T \gg 1/m_\nu$). This discrepancy on the long-range behavior of $V_{\rm bkg}(r)$ is due to the difference between the distributions in Eqs.~(\ref{eq:MBdistribution1}) and (\ref{eq:MBdistribution2}): The former corresponds to the number density of relic neutrinos proportional to $T^3$, while the latter distribution corresponds to the number density that would be exponentially suppressed by $e^{-m_\nu/T}$ for NR neutrinos. In addition, in Ref.~\cite{Ferrer:1999ad}, $V_{\rm bkg}(r)$ was calculated for the standard Fermi-Dirac distribution
\begin{eqnarray}
	\label{eq:FDdistribution1}
	n_{\pm}\left(\mathbf{k},T\right)=\frac{1}{e^{\left(E_k\mp\mu\right)/T}+1}\;,
\end{eqnarray}
for arbitrary chemical potential, but the mass of neutrinos was neglected therein.

However, in the framework of standard cosmology, neutrinos decoupled at $T\sim{\rm MeV}$, after which they were no longer in thermal equilibrium with the cosmic plasma. Instead, they propagated freely until today, maintaining their own distribution:
\begin{eqnarray}
	\label{eq:FDdistribution2}
	n_{\pm}\left(\mathbf{k},T\right)=\frac{1}{e^{\left(\kappa\mp\mu\right)/T}+1}\;.
\end{eqnarray}

The reason why cosmic neutrinos obey the distribution function in Eq.~(\ref{eq:FDdistribution2}), instead of Eq.~(\ref{eq:FDdistribution1}), is that $\kappa$, rather than $E_k$, scales as inversely proportional to the scale factor $a$, i.e., $\kappa\propto 1/a$~\cite{Weinberg2008cosmology}. In the relativistic limit, there is no difference between Eqs.~(\ref{eq:FDdistribution1}) and (\ref{eq:FDdistribution2}). However, we know that the temperature of C$\nu$B today is 
%$T_0=1.95\,{\rm K}=1.68\times 10^{-4}\,{\rm eV}$, 
around $10^{-4}$~eV and neutrino oscillation experiments~\cite{Esteban:2020cvm} tell us that at least two of the three active neutrinos are NR in the C$\nu$B today. Therefore, the results in Refs.~\cite{Ferrer:1998ju,Ferrer:1999ad} using Eqs.~(\ref{eq:MBdistribution2}) and (\ref{eq:FDdistribution1}) only hold for relativistic neutrino background and are \emph{invalid} for the C$\nu$B today, while the computation in Ref.~\cite{Horowitz:1993kw} using Eq.~(\ref{eq:MBdistribution1}) is an approximate result. 

We emphasize that a strict computation of the background effects
on neutrino forces from the C$\nu$B today using Eq.~(\ref{eq:FDdistribution2}) is still lacking, and this is what we do in this section. 

%%%%%%%%%%%%%%
\subsection{Maxwell-Boltzmann distribution}
As a warm-up, we first take the distribution function in Eq.~(\ref{eq:MBdistribution1}), whose massless and NR limits have already been given in Ref.~\cite{Horowitz:1993kw}. 
Substituting
\begin{eqnarray}
	n_+\left(\mathbf{k},T\right)+n_-\left(\mathbf{k},T\right)=2\cosh\left(\frac{\mu}{T}\right)\exp\left(-\frac{\kappa}{T}\right)\;,
\end{eqnarray}
into Eq.~(\ref{eq:V_iso}), we obtain 
\begin{eqnarray}
	\label{eq:VTMBdistribution}
	V_{\rm bkg}(r)=-\frac{G_F^2 g^1_V g^2_V}{2\pi^3}\cosh\left(\frac{\mu}{T}\right)\frac{T}{r^4}\left[\left(1+b^2x^2\right){\cal I}_{\rm MB}\left(x,b\right)-b\frac{\partial}{\partial b}{\cal I}_{\rm MB}\left(x,b\right)\right]C_{\rm L}(T)\;,
\end{eqnarray}
where we have defined the dimensionless quantities
\begin{equation} \label{eq:xby}
x\equiv \frac{m_\nu}{T}, \qquad b\equiv rT, \qquad y\equiv \frac{\kappa}{T}\;,
\end{equation}
with
\begin{equation} 
\label{eq:CL}
C_{\rm L}(T) = \frac{1}{2}\left(1+\frac{p_\nu}{E_\nu+m_\nu}\right),
\end{equation}
and the dimensionless integral   
\begin{eqnarray}
	\label{eq:integralIMB}
	{\cal I}_{\rm MB}(x,b)=\int_0^{\infty}dy \frac{y}{\sqrt{y^2+x^2}}e^{-y}\sin\left(2b y\right)\;.
\end{eqnarray}
The factor $C_{\rm L}(T)$ in Eq.~(\ref{eq:VTMBdistribution}) accounts for the effect of chiral projection of the cosmic background neutrinos into their active component. Here, $E_\nu$ is the average energy and $p_\nu$ is the momentum, and they both depend on the temperature. Note that the modified propagator in Eq.~(\ref{eq:S}) is valid for a general 4-component Dirac spinor. However, in C$\nu$B, only left-handed (LH) chiral neutrinos will contribute to the background force. When cosmic neutrinos are at freeze-out, they are ultra-relativistic and LH helicity state. As the temperature decreases, the mass of the Dirac neutrinos will lead to a population of the right-handed (RH) ones that are sterile to the background force. That is, the factor $C_{\rm L}(T)$ is the amount of LH chiral component in an LH helicity state. It is obvious that $C_{\rm L}=1$ in the massless limit while $C_{\rm L}=1/2$ in the NR limit, which corresponds to the fact that only half of the initial LH helicity neutrinos become LH chiral state when they are non-relativistic.\footnote{We thank Ken Van Tilburg for pointing out this factor of 1/2.}

Eq.~(\ref{eq:integralIMB}) cannot be integrated analytically but can be computed numerically for arbitrary values of $m_\nu$, $T$ and $r$. We are mainly interested in two special scenarios: $x=0$ (the lightest active neutrino can still be massless) and $x\gg1$ (according to the neutrino oscillation experiments, the heaviest active neutrino is at least 0.05 eV, which corresponds to $x\gtrsim 500$ if we consider the temperature of C$\nu$B).

For $x=0$, we have
\begin{eqnarray}
	{\cal I}_{\rm MB}\left(0,b\right)=\frac{2b}{1+4b^2}=\frac{2rT}{1+4r^2T^2}\;,
\end{eqnarray}
and
\begin{eqnarray}
	V_{\rm bkg}(r)=-\frac{8G_F^2  g^1_V g^2_V}{\pi^3}\cosh\left(\frac{\mu}{T}\right)\frac{T^4}{r\left(1+4r^2T^2\right)^2} \qquad (m_\nu=0)\;,
\end{eqnarray}
which is consistent with the result in Refs.~\cite{Horowitz:1993kw,Ferrer:1998ju}. In particular, for high temperatures, $r\gg 1/T$, we notice that $V_T(r)\sim 1/r^5$, which is almost independent of the temperature. For low temperature, $r\ll 1/T$, we find that $V_T(r)\sim T^4/r$.

For $x\gg 1$, since the integral in Eq.~(\ref{eq:integralIMB}) with $y>1$ is exponentially suppressed, the dominant contribution to the integral comes from the region $0<y\ll x$, thus we have
\begin{eqnarray}
	\label{eq:appro1}
	{\cal I}_{\rm MB}\left(x,b\right)\simeq\frac{1}{x}\int_{0}^{\infty}dy y e^{-y}\sin\left(2 b y\right)=\frac{1}{x}\frac{4b}{\left(1+4b^2\right)^2}=\frac{4rT^2}{m\left(1+4r^2T^2\right)^2}\;,
\end{eqnarray}
and
\begin{eqnarray}
	\label{eq:VT1}
	V_{\rm bkg}(r)=-\frac{G_F^2 g^1_V g^2_V}{\pi^3}\cosh\left(\frac{\mu}{T}\right)\frac{m_\nu T^3}{r\left(1+4r^2T^2\right)^2} \qquad (m_\nu\gg T)\;.
\end{eqnarray}
%which is consistent with the result in Ref.~\cite{Horowitz:1993kw}. 
Note that, in contrast to the result in Ref.~\cite{Ferrer:1998ju}, there is no exponential suppression in Eq.~(\ref{eq:VT1}). In particular, for $r\ll 1/T$, we obtain 
\begin{eqnarray}
	V_{\rm bkg}(r)=-\frac{G_F^2  g^1_V g^2_V}{\pi^3}\cosh\left(\frac{\mu}{T}\right)\frac{m_\nu T^3}{r} \qquad (m_\nu \gg T,\;\;r\ll T^{-1})\;,
\end{eqnarray}
while, for $r\gg 1/T$, 
\begin{eqnarray}
	V_{\rm bkg}(r)=-\frac{G_F^2 g^1_V g^2_V}{16\pi^3}\cosh\left(\frac{\mu}{T}\right)\frac{m_\nu}{T}\frac{1}{r^5}\qquad \left(m_\nu\gg T,\;\; r\gg T^{-1}\right)\;,
\end{eqnarray}
which is enhanced by a factor of $m_\nu/T$ compared with the vacuum result in Eq.~(\ref{eq:vacuumpotential}) for NR background neutrinos.

\subsection{Fermi-Dirac distribution}
We now turn to the realistic distribution of background neutrinos in Eq.~(\ref{eq:FDdistribution2}). The first thing to notice is that the  neutrino degeneracy parameter $\zeta\equiv\mu/T$, which characterizes the neutrino–antineutrino asymmetry, is actually very small from constraints of big bang nucleosynthesis: $\zeta\lesssim {\cal O}\left(10^{-2}\right)$~\cite{Serpico:2005bc,Lesgourgues:2006nd}. Therefore, we can expand the neutrino distribution function into a series of $\zeta$,
\begin{eqnarray}
	n_+\left(\mathbf{\kappa},T\right)+n_-\left(\mathbf{\kappa},T\right)=\frac{2}{e^{\kappa/T}+1}+{\cal O}\left(\zeta^2\right)\;,
\end{eqnarray}
and only take the leading-order term, which is independent of $\zeta$. Then the background potential turns out to be
\begin{eqnarray}
	\label{eq:VTFDdistribution}
	V_{\rm bkg}(r)=-\frac{G_F^2 g^1_V g^2_V}{2\pi^3}\frac{T}{r^4}\left[\left(1+b^2x^2\right){\cal I}_{\rm FD} \left(x,b\right)-b\frac{\partial}{\partial b}	{\cal I}_{\rm FD}\left(x,b\right)\right]C_{\rm L}(T)\;,
\end{eqnarray} 
where $x$, $b$, and $y$ are defined in Eq.~\eqref{eq:xby}  and
\begin{eqnarray}
	\label{eq:integralIFD}
	{\cal I}_{\rm FD}\left(x,b\right)=\int_{0}^{\infty}dy \frac{y}{\sqrt{y^2+x^2}}\frac{1}{e^y+1}\sin\left(2b \,y\right)\;.
\end{eqnarray}
The integral in Eq.~(\ref{eq:integralIFD}) can be numerically calculated for arbitrary values of $m_\nu$, $T$, and $r$. In the massless limit $(x=0)$ and NR limit $(x\gg 1)$, ${\cal I}_{\rm FD}\left(x,b\right)$ can be carried out analytically.

For $x=0$, we have
\begin{eqnarray}
	{\cal I}_{\rm FD}\left(x,b\right)=\frac{1}{4}\left[\frac{1}{b}-2\pi\,{\rm csch}\left(2\pi b\right)\right]\;,
\end{eqnarray}
and the background potential
\begin{eqnarray}
	V_{\rm bkg}(r)=-\frac{G_F^2  g^1_V g^2_V}{4\pi^3}\frac{1}{r^5}\left\{1-\pi r T{\rm csch}\left(2\pi r T\right)\left[1+2\pi rT {\rm coth}\left(2\pi r T\right)\right]\right\}\quad (m_\nu=0)\;,
		\label{eq:FDmassless}
\end{eqnarray}
which is consistent with the result obtained in Ref.~\cite{Ferrer:1999ad}, where the neutrino distribution Eq.~(\ref{eq:FDdistribution1}) was taken but the neutrino mass was neglected. An interesting observation is that, in the long-range limit,
\begin{eqnarray}
	V_{\rm bkg}(r)=-\frac{G_F^2 g^1_V g^2_V}{4\pi^3}\frac{1}{r^5}\qquad (m_\nu =0,\;\;r\gg T^{-1})\;,
\end{eqnarray}
which happens to be the opposite of Eq.~(\ref{eq:vacuumpotential}). This means that, for massless neutrinos in the limit $\zeta \to 0$, the vacuum potential is completely screened off by the C$\nu$B.

%%%%%%%%%%%%%%%%%%%%%%%%%%%% Table 2 %%%%%%%%%%%%%%%%%%%%%%%%%%
\renewcommand\arraystretch{1.4}
\begin{table}[t!]
	\centering
	\begin{tabular}{c|c|c|c|c}
		\hline\hline
		$\nu$BDF & $m_\nu=0$, $r\ll T^{-1}$ & $m_\nu=0$, $r\gg T^{-1}$ & $m_\nu\gg T$, $r\ll T^{-1}$ & $m_\nu\gg T$, $r\gg T^{-1}$\\
		\hline
		MB & $-\frac{8}{\pi^3}G_F^2 g^1_V g^2_V\frac{T^4}{r}$  & $-\frac{1}{2\pi^3}G_F^2 g^1_V g^2_V \frac{1}{r^5}$ &$-\frac{1}{\pi^3}G_F^2 g^1_V g^2_V \frac{m_\nu T^3}{r}$&$-\frac{1}{16\pi^3}G_F^2 g^1_V g^2_V\frac{m_\nu}{T}\frac{1}{r^5}$\\
		\hline
		FD &$-\frac{7\pi}{90}G_F^2 g^1_V g^2_V\frac{T^4}{r}$&$-\frac{1}{4\pi^3}G_F^2 g^1_V g^2_V \frac{1}{r^5}$&$-\frac{3\zeta(3)}{4\pi^3}G_F^2 g^1_V g^2_V \frac{m_\nu T^3}{r}$&$-\frac{1}{64\pi^3}G_F^2 g^1_V g^2_V\frac{m_\nu}{T}\frac{1}{r^5}$\\
		\hline\hline
	\end{tabular}
	\vspace{0.5cm}
	\caption{\label{table:comparison}Comparison of the short- and long-range behaviors of the background potential $V_{\rm bkg}(r)$ in the massless and non-relativistic limits with the neutrino Background Distribution Function ($\nu$BDF) taking the Maxwell-Boltzmann (MB) distribution in Eq.~(\ref{eq:MBdistribution1}) and Fermi-Dirac (FD) distribution in Eq.~(\ref{eq:FDdistribution2}). We have neglected the chemical potential in both distribution functions.}
\end{table}	
\renewcommand\arraystretch{1.0}
%%%%%%%%%%%%%%%%%%%%%%%%%%%%%%%%%%%%%%%%%%%%%%%%%%%%%%%%%%%%%%%%

Let us now take a look at the NR limit of Eq.~(\ref{eq:integralIFD}). As with the case of Boltzmann distribution, for $x\gg1$, one obtains
\begin{eqnarray}
	\label{eq:appro2}
	{\cal I}_{\rm FD}\left(x,b\right)&\simeq& \frac{1}{x}\int_{0}^{\infty}dy \frac{y}{e^y+1}\sin\left(2b y\right)\nonumber\\
	&=&\frac{i}{8x}\left[\psi^{(1)}\left(\frac{1}{2}+ib\right)-\psi^{(1)}\left(\frac{1}{2}-ib\right)+\psi^{(1)}\left(1-ib\right)-\psi^{(1)}\left(1+ib\right)\right],
\end{eqnarray}  
where the $n$-th ordered polygamma function is defined as
\begin{equation}
\psi^{(n)}(z)= \frac{d}{dz}\psi^{(n-1)}(z)=\frac{d^{n+1}}{dz^{n+1}} \log \Gamma(z)\;,
\end{equation}
with $\Gamma(z)$ being the gamma function. Therefore, the background potential of NR cosmic neutrinos turns out to be
\begin{eqnarray}
\label{eq:NRexpressionDirac}
V_{\rm bkg}(r)&=&-i\frac{G_F^2 g^1_V g^2_V}{32\pi^3}\frac{T^2}{mr^4}\left\{\left[\psi^{(1)}\left(\frac{1}{2}+ib\right)-\psi^{(1)}\left(\frac{1}{2}-ib\right)+\psi^{(1)}\left(1-ib\right)-\psi^{(1)}\left(1+ib\right)\right]\right.\nonumber\\
&\times &\left.\left(1+b^2x^2\right)-ib\left[\psi^{(2)}\left(\frac{1}{2}+ib\right)+\psi^{(2)}\left(\frac{1}{2}-ib\right)-\psi^{(2)}\left(1-ib\right)-\psi^{(2)}\left(1+ib\right)\right]
\right\},\nonumber\\
&&\qquad\qquad\qquad\qquad\qquad\qquad\qquad\qquad\qquad\qquad\qquad\qquad\qquad (m_\nu\gg T)
\end{eqnarray}
In particular, for $r\ll 1/T$, i.e., $b\ll 1$, we have
\begin{eqnarray}
V_{\rm bkg}(r)&=&-\left[\psi^{(2)}(1)-\psi^{(2)}\left(\frac{1}{2}\right)\right]\frac{G_F^2 g^1_V g_V^2}{16\pi^3}\frac{m_\nu T^3}{r}\nonumber\\
&=&-\frac{3\zeta(3)}{4\pi^3}G_F^2 g^1_V g^2_V\frac{m_\nu T^3}{r}\qquad \left(m_\nu\gg T,\;\;r\ll T^{-1}\right),
\end{eqnarray}
with $\zeta(3)\simeq 1.202$ the Riemann zeta function,
while, for the long-range limit $b\gg 1$, one obtains
\begin{eqnarray}
V_{\rm bkg}(r)=-\frac{G_F^2 g^1_V g^2_V}{64\pi^3}\frac{m_\nu}{T}\frac{1}{r^5}\qquad \left(m_\nu\gg T,\;\;r\gg T^{-1}\right),
\end{eqnarray}
which is, as in the case of the Boltzmann distribution, enhanced by a factor of $m_\nu/T$ compared with the vacuum potential in Eq.~(\ref{eq:vacuumpotential}). 

To sum up, we have provided in Eq.~(\ref{eq:VTFDdistribution}) the general background potential valid for any temperatures and distances and discussed the special scenarios in the massless and NR neutrinos limits, which have simple analytical expressions. Compared to the results of Maxwell-Boltzmann distribution in last subsection, we conclude that both distributions lead to similar short-range and long-range behaviors of the background potential  in the massless limit ($m_\nu=0$) and NR limit ($m_\nu \gg T$), up to some numerical factors (cf. Table~\ref{table:comparison}).

\subsection{The case of Majorana neutrinos}
\label{subsec:majorana case}

The above calculations for Dirac neutrinos can be generalized to the scenario of Majorana neutrinos. If $\nu$ is a Majorana neutrino with mass $m_\nu$, then its general four-fermion interaction is given by  
\begin{eqnarray}
\label{eq:lagMajorama}
{\cal L}_{\rm int}=\frac{G_F}{\sqrt{2}}\left[\bar{\nu}\gamma^\mu\gamma_5\nu\right]\left[\bar{\chi}\gamma_\mu\left(g_V^{\chi}+g_A^{\chi}\gamma_5\right)\chi\right]\;,
\end{eqnarray}
where we have used the identity $\bar{\nu}\gamma^\mu \nu=0$ for Majorana fermions comparing with Eq.~(\ref{eq:lagDirac}). Taking into account the modified neutrino propagator due to the background,  Eq.~(\ref{eq:S}), the scattering amplitude reads
\begin{eqnarray}
i{\cal A}(q)=\frac{G_{F}^{2} g^1_V g^2_V}{2}\int\frac{d^{4}k}{\left(2\pi\right)^{4}}{\rm Tr}\left[\gamma^{0}\gamma_5S_{\nu}(k)\gamma^{0}\gamma_5S_{\nu}\left(k+q\right)\right]\times 2\;,
\label{eq:amplitudeMajorana}
\end{eqnarray}
where the factor of 2 is due to the exchange of two neutrino propagators in the loop. As with the Dirac case, the background effect comes from the crossed terms. After some algebra, one obtains
\begin{eqnarray}
{\cal A}_{{\rm bkg}}(\mathbf{q})=4G_{F}^{2} g^1_V g^2_V\int\frac{d^{3}\mathbf{k}}{\left(2\pi\right)^{3}}\frac{n_{+}\left(\mathbf{k}\right)+n_{-}\left(\mathbf{k}\right)}{2E_{k}}\left[\frac{2\left|\mathbf{k}\right|^{2}+\mathbf{k}\cdot\mathbf{q}}{2\mathbf{k}\cdot\mathbf{q}+\left|\mathbf{q}\right|^{2}}+\left(\mathbf{k}\to-\mathbf{k}\right)\right].\label{eq:AbkgMajorana}
\end{eqnarray}
For isotropic distributions $n_{\pm}(\mathbf{k})=n_{\pm}(\kappa)$, Eq.~(\ref{eq:AbkgMajorana}) can be reduced to
\begin{eqnarray}
V_{{\rm bkg}}(r)=-\frac{G_{F}^{2} g^1_V g^2_V}{4\pi^{3}r^{4}}\int_{0}^{\infty}d\kappa\, \kappa\,\frac{n_{+}\left(\kappa\right)+n_{-}\left(\kappa\right)}{\sqrt{\kappa^{2}+m_{\nu}^{2}}}\left[\sin\left(2\kappa r\right)-2\kappa r\cos\left(2\kappa r\right)\right]\;,\label{eq:V_isoMajorana}
\end{eqnarray} 
which, as expected, matches the result for Dirac neutrinos in Eq.~(\ref{eq:V_iso}) in the massless limit.

We then take the Fermi-Dirac distribution in Eq.~(\ref{eq:FDdistribution2}) to calculate $V_{{\rm bkg}}(r)$ in the C$\nu$B. Note that for Majorana neutrinos, the chemical potential vanishes, so that 
\begin{equation}
n_+(\kappa)=n_-(\kappa)=\frac{1}{e^{\kappa/T}+1}\;.
\end{equation}
Therefore, the background potential turns out to be
\begin{eqnarray}
\label{eq:VTFDdistributionMajorana}
V_{\rm bkg}(r)=-\frac{G_F^2 g^1_V g^2_V}{2\pi^3}\frac{T}{r^4}\left[{\cal I}_{\rm FD} \left(x,b\right)-b\frac{\partial}{\partial b}	{\cal I}_{\rm FD}\left(x,b\right)\right]\;,
\end{eqnarray} 
where $x$, $b$, and the integral ${\cal I}_{\rm FD}^{}(x,b)$ is defined in Eqs.~(\ref{eq:xby}) and (\ref{eq:integralIFD}). Note that for Majorana neutrinos, there is no need to include the factor of chiral projection (\ref{eq:CL}) as the Dirac neutrino case.

In the massless limit ($x=0$), it is obvious that $V_{\rm bkg}(r)$ is the same as we have for the Dirac neutrino case in Eq.~(\ref{eq:FDmassless}). 

In the NR limit ($x\gg 1$), ${\cal I}_{\rm FD}^{}(x,b)$ can be integrated analytically and is given by Eq.~(\ref{eq:appro2}). Therefore the background potential turns out to be
\begin{eqnarray}
\label{eq:NRexpressionMajorana}
V_{\rm bkg}(r)&=&-i\frac{G_F^2 g^1_V g^2_V}{16\pi^3}\frac{T^2}{mr^4}\left\{\left[\psi^{(1)}\left(\frac{1}{2}+ib\right)-\psi^{(1)}\left(\frac{1}{2}-ib\right)+\psi^{(1)}\left(1-ib\right)-\psi^{(1)}\left(1+ib\right)\right]\right.\nonumber\\
&&\left.-ib\left[\psi^{(2)}\left(\frac{1}{2}+ib\right)+\psi^{(2)}\left(\frac{1}{2}-ib\right)-\psi^{(2)}\left(1-ib\right)-\psi^{(2)}\left(1+ib\right)\right]
\right\}\;.
\end{eqnarray}
In particular, for the short-range limit ($b\ll 1$) one obtains
\begin{eqnarray}
V_{\rm bkg}(r)&=&-\left[\psi^{(4)}(1)-\psi^{(4)}\left(\frac{1}{2}\right)\right]\frac{G_F^2 g^1_V g^2_V}{24\pi^3}\frac{T^5}{m_\nu r}\nonumber\\
&=&-\frac{30\zeta(5)}{\pi^3}G_F^2 g^1_V g^2_V\frac{T^5}{m_\nu r}
\qquad \left(m_\nu\gg T,\;\;r\ll T^{-1}\right)\;,
\end{eqnarray}
with $\zeta(5)\simeq 1.037$, while for the long-range limit ($b\gg 1$), we have 
\begin{eqnarray}
V_{\rm bkg}(r)=-\frac{G_F^2  g^1_V g^2_V}{8\pi^3}\frac{1}{m_\nu T r^7}\qquad \left(m_\nu\gg T,\;\; r\gg T^{-1}\right)\;.
\end{eqnarray}

%%%%%%%%%%%%%%%%%%%%%%%%%%%% Table 3 %%%%%%%%%%%%%%%%%%%%%%%%%%
\renewcommand\arraystretch{1.4}
\begin{table}[t!]
\centering
\begin{tabular}{c|c|c|c}
\hline\hline
nature of neutrino & general expression & $r\ll T^{-1}$ & $r\gg T^{-1}$ \\
\hline
Dirac & Eq.~(\ref{eq:NRexpressionDirac})  & $-\frac{3\zeta(3)}{4\pi^3}G_F^2 g^1_V g^2_V\frac{m_\nu T^3}{r}$ &$-\frac{1}{64\pi^3}G_F^2 g^1_V g^2_V\frac{m_\nu}{T}\frac{1}{r^5}$\\
\hline
Majorana & Eq.~(\ref{eq:NRexpressionMajorana}) &$-\frac{30\zeta(5)}{\pi^3}G_F^2 g^1_V g^2_V\frac{T^5}{m_\nu r}$&$-\frac{1}{8\pi^3}G_F^2 g^1_V g^2_V\frac{1}{m_\nu T}\frac{1}{r^7}$\\
\hline\hline
\end{tabular}
\vspace{0.5cm}
\caption{\label{table:comparisonDiracMajorana}Comparison of the short- and long-range behavior of the background potential $V_{\rm bkg}(r)$ in non-relativistic C$\nu$B ($m_\nu\gg T$) with $n_{\pm}$ taking the Fermi-Dirac distribution in Eq.~(\ref{eq:FDdistribution2}) for Dirac and Majorana background neutrinos.}
\end{table}	
\renewcommand\arraystretch{1.0}
%%%%%%%%%%%%%%%%%%%%%%%%%%%%%%%%%%%%%%%%%%%%%%%%%%%%%%%%%%%%%%%%

In Table~\ref{table:comparisonDiracMajorana}, we have compared the short- and long-range behaviors of the background potential $V_{\rm bkg}(r)$ due to Dirac and Majorana neutrinos in the NR regime. Notice that, at short distances ($r\ll T^{-1}$ and $m_\nu^{-1} \ll T^{-1}$), the background potential of Majorana neutrinos differs from that of Dirac neutrinos by a factor of $m_\nu^2/T^2\gg 1$. Whereas, at long distances ($r\gg T^{-1}\gg m_\nu^{-1}$), the relative factor is $m_\nu^2r^2\gg 1$. This difference can be understood by the fact that the mass term in the neutrino propagator dominates in the NR limit, and there should be two mass insertions in the Dirac-neutrino propagator compared to just one mass insertion in the Majorana-neutrino propagator. Therefore, we conclude that for NR background neutrinos, the background potential of Dirac neutrinos is much larger than that of Majorana neutrinos at both long and short distances.

\subsection{Discussion}
We close this section by briefly summarizing the main results of the thermal corrections to neutrino forces from cosmic background neutrinos.

Neutrinos in the C$\nu$B are NR today (although the lightest neutrino can still be massless) and obey the Fermi-Dirac distribution in Eq.~(\ref{eq:FDdistribution2}) with negligible chemical potential. The general expressions of the finite-temperature corrections, valid for arbitrary neutrino masses and distances, are given by Eqs.~(\ref{eq:VTFDdistribution}) and (\ref{eq:VTFDdistributionMajorana}) for Dirac and Majorana neutrinos, respectively. In the massless limit, the background potential $V_{\rm bkg}(r)$ is the same for Dirac and Majorana neutrinos. However, for NR background neutrinos, $V_{\rm bkg}(r)$ is much larger for Dirac neutrinos. This distinction can, at least in principle, be used to determine the nature of neutrinos.

The most remarkable feature of the background potential from C$\nu$B is that, at large distances ($r\gg 1/m_\nu$), it is \emph{not} exponentially suppressed, whereas the vacuum potential is suppressed by $e^{-2m_\nu r}$~\cite{Grifols:1996fk}. This is because the number density of background neutrinos in the C$\nu$B is always proportional to $T^3$, no matter whether they are relativistic or not. Since the total potential is given by adding the vacuum part and the background part, neutrino forces between two objects will be dominated by the corrections of C$\nu$B in the long-range limit for massive mediated neutrinos. However, neutrino forces including thermal corrections of C$\nu$B are still too small to reach the experimental sensitivities today (cf. Sec.~\ref{sec:exp}). Below we will discuss neutrino forces in other higher-energy neutrino backgrounds, which might offer prospects of experimental detection in the near future.

Finally, we  comment on the controversial topic of many-body neutrino forces in neutron stars. In Ref.~\cite{Fischbach:1996qf}, a catastrophically large many-body neutrino force was obtained using the vacuum neutrino propagator. 
Matter effects due to the neutrons have been computed in Ref.~\cite{Notzold:1987ik}. It was claimed in ~\cite{Smirnov:1996vj,Abada:1996nx,Kachelriess:1997cr,Kiers:1997ty,Abada:1998ti,Arafune:1998ft} that this changes the result of Ref.~\cite{Fischbach:1996qf}. Our result is irrelevant to this issue as we only consider the neutrino background, and we do not elaborate any further.

%{\color{blue}Finally, we would like to comment on a controversial effect of many-body neutrino forces in neutron  stars~\cite{Fischbach:1996qf,Smirnov:1996vj,Abada:1996nx,Kachelriess:1997cr,Kiers:1997ty,Abada:1998ti,Arafune:1998ft}. In an ideal neutron star without on-shell neutrino backgrounds, the neutrino propagator can still be modified due to the dense neutrons contributing at the loop level. This leads to a neutron-modified neutrino propagator, which has been computed in Ref.~\cite{Notzold:1987ik}. The catastrophically large many-body neutrino force proposed in Ref.~\cite{Fischbach:1996qf} was obtained using the vacuum neutrino propagator. It might be possible that with the modified neutrino propagator, the effect could drastically reduced, though a definite answer would a require dedicated calculation which is beyond the scope of this work. }

\section{Neutrino forces with directional neutrino backgrounds}
\label{sec:reactorsolor}

In this section, we move to discuss anisotropic backgrounds. In particular, we consider one with a specific direction.
Reactor, solar, and supernova neutrinos are example for such cases.

\subsection{Calculations}
Reactor, solar, and supernova neutrinos are anisotropic and much more energetic than cosmic relic
neutrinos. Solar neutrinos arrive at the Earth with an almost certain direction. Reactor neutrinos can also be assumed to travel in a fixed direction if the sizes of the reactor core and the detector are much smaller than the distance between them. 
In addition, we also consider a galactic (10 kpc) supernova neutrino burst. 
Although such an event is rare ($2\sim3$ times per century), its neutrino flux is orders of magnitude higher than solar neutrinos with an extremely small angular spread, providing a unique opportunity for future experiments to search for such forces.   

In order to compute the effect of these backgrounds on the neutrino force, we make two well-motivated assumptions:
\begin{enumerate}
\item We assume that the neutrino flux has a directional distribution with all neutrinos moving in the same direction. 
For solar and supernova neutrinos, this is a good approximation, whereas for reactor neutrinos it requires that the size of the reactor core and detector are much smaller than the distance between them.
\item We assume that the neutrino flux is monochromatic, i.e., all neutrinos in flux have the same energy. Although this is not exactly true, it is worth mentioning that among the four well-measured
solar neutrino spectra ($^{8}$B, $^{7}$Be, $pep$, $pp$), two of
them ($^{7}$Be, $pep$) are indeed monochromatic.
\end{enumerate}
With these assumptions of directionality and monochromaticity, we consider the following distribution:
\begin{equation}
n_{\pm}\left(\mathbf{k}\right)=\left(2\pi\right)^{3}\delta^{3}\left(\mathbf{k}-\mathbf{k}_{0}\right)\Phi_{0}\thinspace,\label{eq:x-1}
\end{equation}
where $\Phi_{0}=\int n_{\pm}\left(\mathbf{k}\right)d^{3}\mathbf{k}/\left(2\pi\right)^{3}$
is the flux of neutrinos.  
Although actual reactor and solar neutrino spectra are not monochromatic, our result derived below based on Eq.~\eqref{eq:x-1} can be applied
to a generic spectrum by further integrating over $\mathbf{k}_{0}$,
weighted by the corresponding $\Phi_{0}$, since any spectrum can be
expressed as a superposition of delta functions. For the treatment of a directional spectrum with a finite energy spread, see Appendix~\ref{app:spread}.

%%%%%%%%%%%%%%%%%%%%%%%%%%%%%%% Fig.2 %%%%%%%%%%%%%%%%%%%%%%%%%%%%%%%
\begin{figure}[t!]
	\centering 	
	\includegraphics[width=7cm]{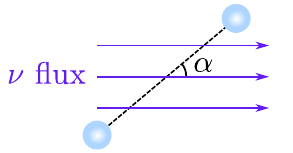} \caption{\label{fig:alpha}An illustration of neutrino forces between two objects in a directional neutrino flux background.}
\end{figure}
%%%%%%%%%%%%%%%%%%%%%%%%%%%%%%%%%%%%%%%%%%%%%%%%%%%%%%%%%%%%%%%%%%%%

The anisotropic background leads to an anisotropic scattering amplitude,
and hence an anisotropic potential that depends not only on $r$
but also on the angle between $\mathbf{k}_{0}$ and $\mathbf{r}$,
denoted by $\alpha$ (cf. Fig.~\ref{fig:alpha}). Without loss of generality, we assume $\mathbf{k}_{0}$
is aligned with the $z$-axis and $\mathbf{r}$ lies in the $x$-$z$ plane:
\begin{eqnarray}
\mathbf{k}_{0}=E_{\nu}\left(0,0,1\right),\quad\mathbf{r}=r\left(s_{\alpha},0,c_{\alpha}\right), \label{eq:alpha}
\end{eqnarray}
where $(c_{\alpha},\ s_{\alpha})\equiv\left(\cos\alpha,\ \sin\alpha\right)$.

Substituting the distribution \eqref{eq:x-1} into Eq.~(\ref{eq:Abkg}), we obtain

\begin{align}
{\cal A}_{{\rm bkg}}\left(\mathbf{q}\right) & =2G_{F}^{2}g_{V}^{1}g_{V}^{2}\frac{\Phi_{0}}{E_{\nu}}\left[\frac{2E_{\nu}^{2}+\mathbf{k}_{0}\cdot\mathbf{q}}{\rho^{2}+2\mathbf{k}_{0}\cdot\mathbf{q}}+\frac{2E_{\nu}^{2}-\mathbf{k}_{0}\cdot\mathbf{q}}{\rho^{2}-2\mathbf{k}_{0}\cdot\mathbf{q}}\right]\nonumber \\
& =8G_{F}^{2}g_{V}^{1}g_{V}^{2}\Phi_{0}E_{\nu}\frac{1-\xi^{2}}{\rho^{2}-4E_{\nu}^{2}\xi^{2}},\label{eq:1011}
\end{align}
where $\rho\equiv\left|\mathbf{q}\right|$ and 
\begin{equation}
\xi\equiv\frac{\mathbf{k}_{0}\cdot\mathbf{q}}{|\mathbf{k}_{0}||\mathbf{q}|}\thinspace. \label{eq:defxi}
\end{equation}
Note that the typical energy of reactor and solar neutrinos is ${\cal O}$(MeV), so we can safely neglect the neutrino mass in Eq.~(\ref{eq:Abkg}).
Thus, the background-induced potential  is given by
\begin{eqnarray}
\label{eq:anisotropicpotential}
V_{{\rm bkg}}(\mathbf{r})=-\int\frac{d^{3}\mathbf{q}}{\left(2\pi\right)^{3}}e^{i\mathbf{q}\cdot\mathbf{r}}{\cal A}_{{\rm bkg}}(\mathbf{q})=-\frac{g^1_V g^2_V}{\pi^{3}}G_{F}^{2}\Phi_{0}E_{\nu}^{2}\times{\cal I}\;,
\end{eqnarray}
where ${\cal I}$ is a dimensionless integral. We further define
\begin{equation}
\ell\equiv rE_{\nu}
\end{equation}
and note that ${\cal I}$ is depends only on $\ell$
and $\alpha$:
\begin{equation}
{\cal I}(\ell,\alpha)\equiv\frac{1}{E_{\nu}}\int d^{3}\mathbf{q}e^{i\mathbf{q}\cdot\mathbf{r}}\frac{1-\xi^{2}}{\rho^{2}-4E_{\nu}^{2}\xi^{2}}\thinspace.\label{eq:I_def}
\end{equation}
%%%%%%%%%%%%%%%%%%%%%%%%%%% Fig.3 %%%%%%%%%%%%%%%%%%%%%%%%%%%%%%%%%
\begin{figure}[t!]
\centering 
	
\includegraphics[width=0.85\textwidth]{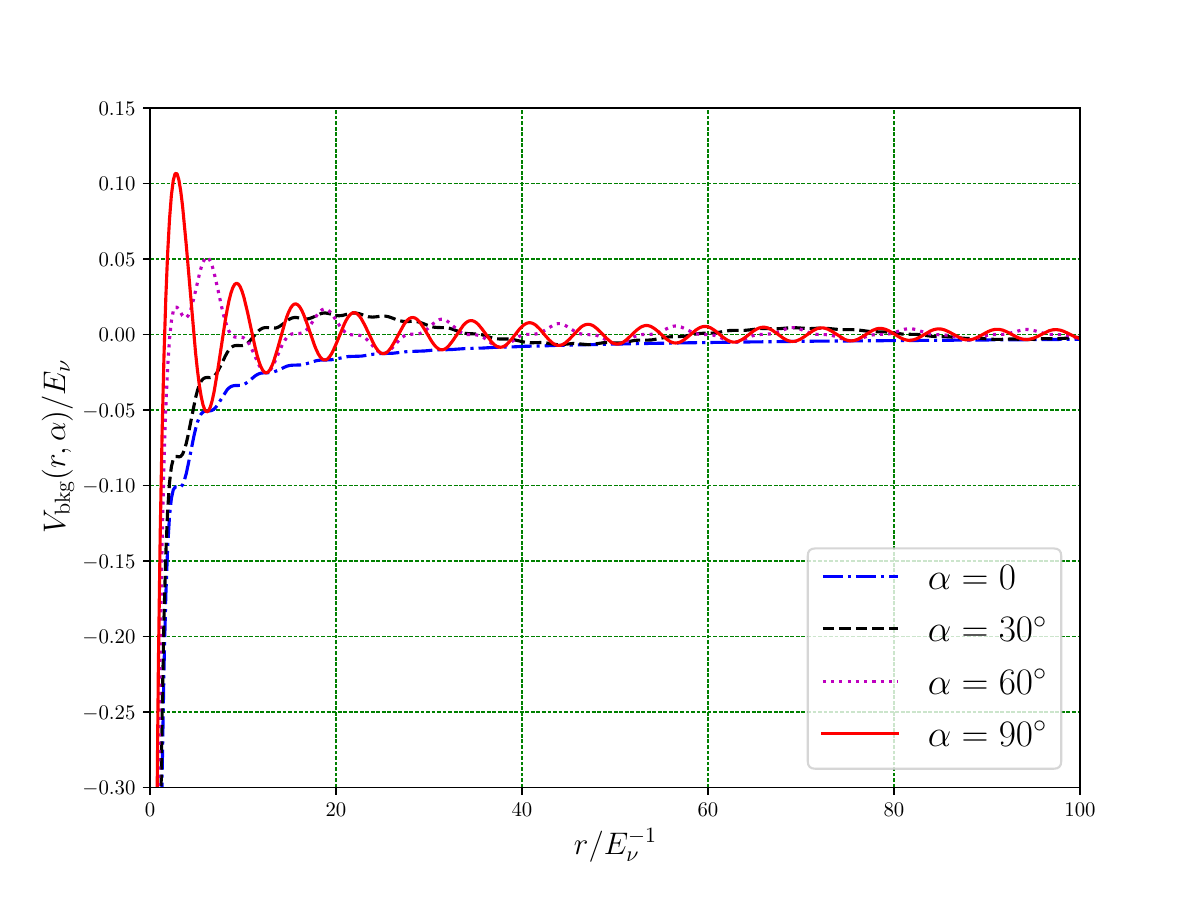} \caption{\label{fig:int} Evolution of the directional background potential with the distance for $\alpha=0$, $30^{\circ}$, $60^{\circ}$ and $90^{\circ}$. Notice that the distance $r$ is in the unit of $E_{\nu}^{-1}$ while the background potential $V_{\rm bkg}(r,\alpha)$ is in the unit of $E_\nu$. In addition, an overall dimensionless factor, $G_{F}^{2}g_{V}^{1}g_{V}^{2}\Phi_0 E_{\nu}$,
has been omitted for the background potential.}
\end{figure}
%%%%%%%%%%%%%%%%%%%%%%%%%%%%%%%%%%%%%%%%%%%%%%%%%%%%%%%%%%%%%%%%%%
%
In Appendix~\ref{sec:integral}, we show that, for generic $\alpha$
and $\ell$, the integral can be reduced to
\begin{align}
{\cal I}\left(\ell,\alpha\right) =\frac{\pi^{2}}{2\ell}\left(3+\cos2\alpha\right)
-2\pi\int_{-1}^{1}d\xi\,\xi\left(1-\xi^{2}\right)\int_{0}^{\pi}d\varphi\,\sin\left(2\ell\xi\left|c_{\alpha}\xi+s_{\alpha}\sqrt{1-\xi^{2}}\cos\varphi\right|\right).\label{eq:I_full}
\end{align}
For the special cases of $\alpha=0$ and $\alpha=\pi/2$, we find
\begin{align}
{\cal I}\left(\ell,\alpha=0\right) & =\frac{\pi^{2}}{\ell}\left[1+\frac{\sin2\ell}{2\ell}\right]\;, \label{eq:I-0}\\
{\cal I}\left(\ell,\alpha=\frac{\pi}{2}\right) & =\frac{\pi^{2}}{\ell}\left[1-4\ell\int_{0}^{1}d\xi\xi\left(1-\xi^{2}\right)H_{0}\left(2\ell\xi\sqrt{1-\xi^{2}}\right)\right]\;,
\end{align}
where $H_{0}$ is the zeroth-order Struve H function.\footnote{
We note that  {\tt Mathematica} contains some unidentified bug leading to incorrect results of integrals involving the Struve H function, 
e.g. $\int_0^1 H_0(\sqrt{1-z^2} z) \, dz$ should be nonzero while {\tt Integrate} in  {\tt Mathematica}  only produces a vanishing result. The bug has been confirmed by the developers of  {\tt Mathematica}.}  For generic values of $\alpha$, though we cannot carry out the
integration analytically, Eq.~\eqref{eq:I_full} can be readily used
to compute ${\cal I}\left(\ell,\alpha\right)$ numerically. We have numerically
verified that $\int{\cal I}\left(\ell,\alpha\right)dc_{\alpha}$ can reproduce the $r^{-4}$ dependence in Eq.~\eqref{eq:V_iso}, which is expected when Eq.~\eqref{eq:V_iso} is applied to an isotropic and monochromatic flux. For illustration, in Fig.~\ref{fig:int} we show the evolution of the directional background potential $V_{\rm bkg}$ with the distance $r$ for $\alpha=0$, $\pi/6$, $\pi/3$ and $\pi/2$.

%%%%%%%%%%%%%%%%%%%%%%%%%%%%%%% Fig.4 %%%%%%%%%%%%%%%%%%%%%%%%%%%%%
\begin{figure}[t!]
\centering
\includegraphics[width=0.49\textwidth]{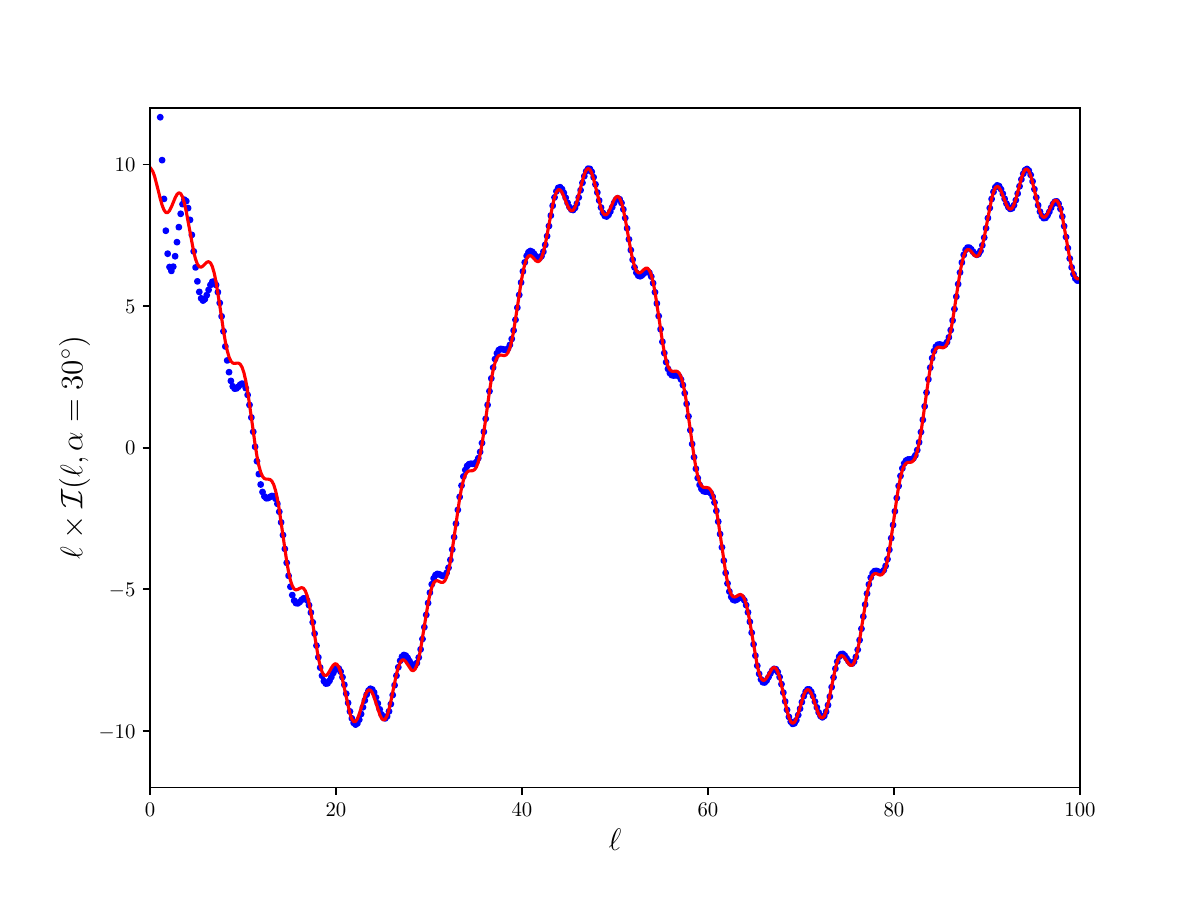}
\includegraphics[width=0.49\textwidth]{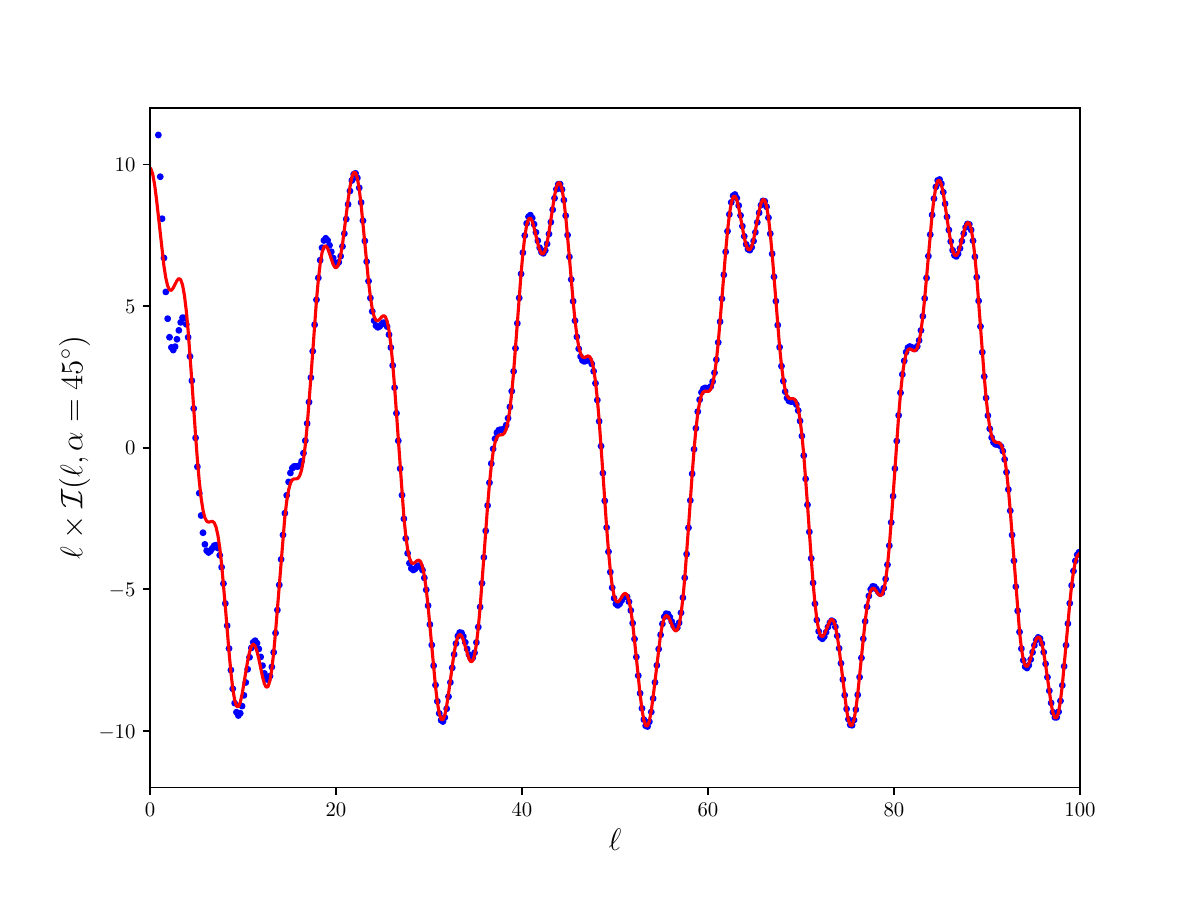}\\
\includegraphics[width=0.49\textwidth]{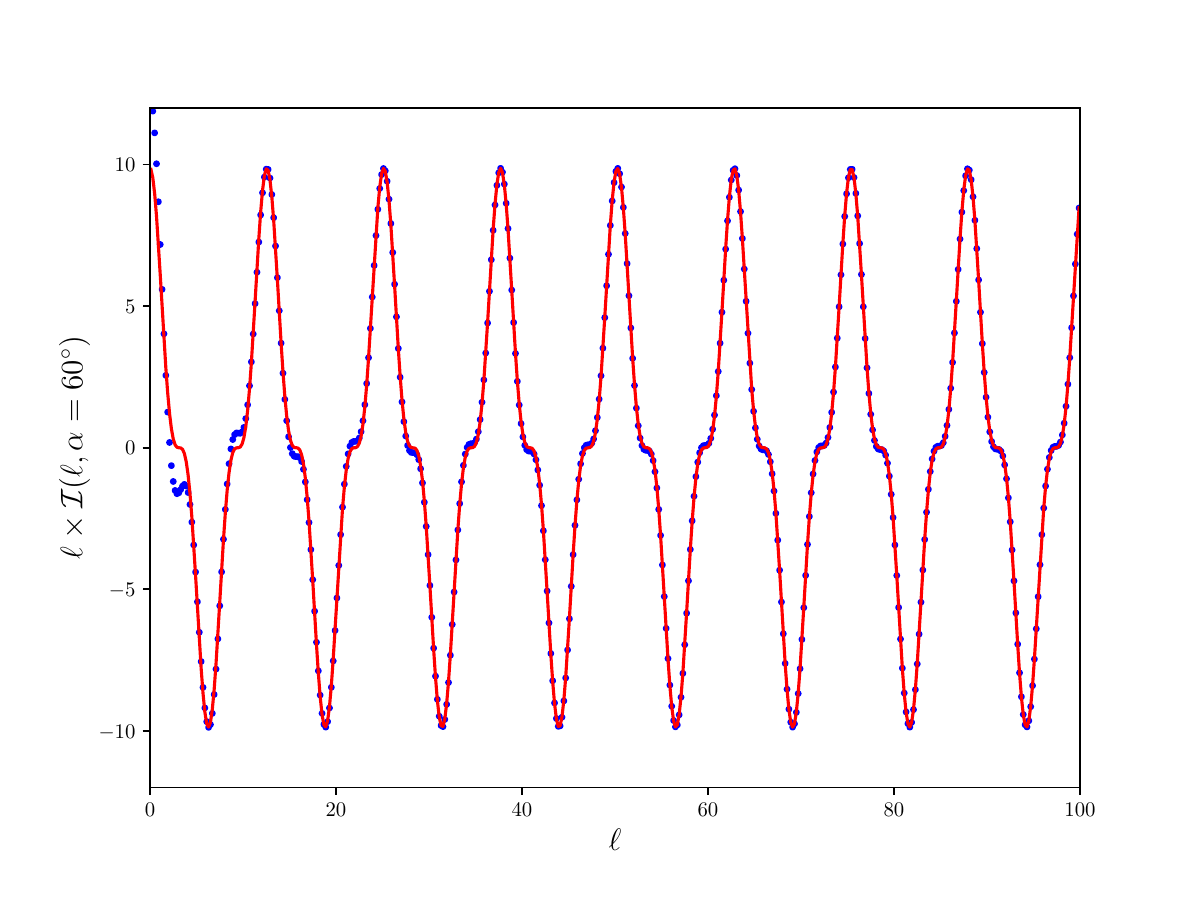}
\includegraphics[width=0.49\textwidth]{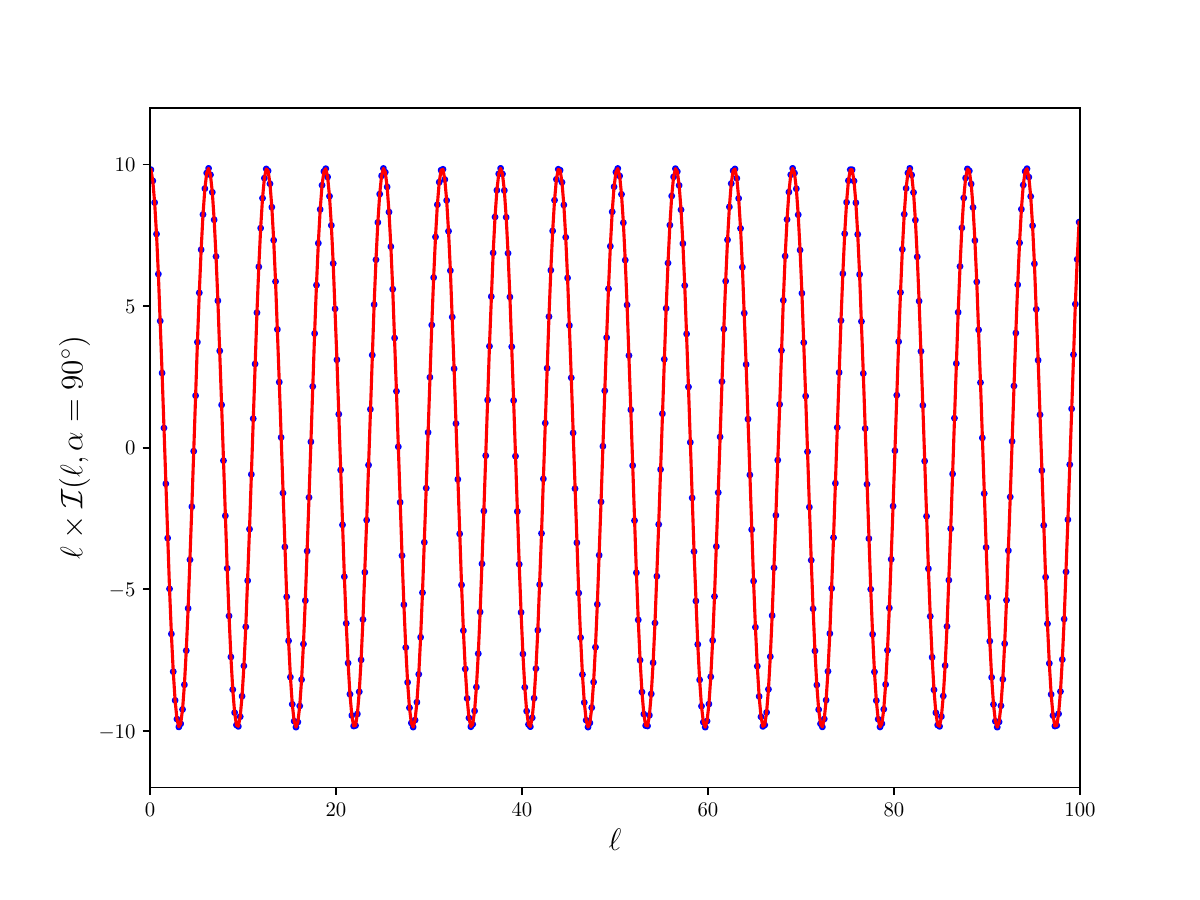}
\caption{\label{fig:compare} Comparison between the numerical results of $\ell\times{\cal I}(\ell ,\alpha)$ computed from Eq.~(\ref{eq:I_full}) (blue dotted points) and the analytical results computed from Eq.~(\ref{eq:analytical}) (red solid line) for $\alpha=30^{\circ}$, $\alpha=45^{\circ}$, $\alpha=60^{\circ}$ and $\alpha=90^{\circ}$. They match excellently at large distances (i.e., $r \gg E_\nu^{-1}$).}
\end{figure}
%%%%%%%%%%%%%%%%%%%%%%%%%%%%%%%%%%%%%%%%%%%%%%%%%%%%%%%%%%%%%%%%%%%%%%

At long distances ($\ell \gg 1$), the numerical evaluation of the double integral in Eq.~\eqref{eq:I_full} is computationally expensive. We find that ${\cal I}(\ell,\alpha)$ has a simple analytical expression for $\ell \gg 1$:
\begin{eqnarray}
	\label{eq:analytical}
{\cal I}\left(\ell\gg 1,\alpha\right)=\frac{\pi^2}{\ell} \cos^2\left(\frac{\alpha}{2}\right)\cos\left[\left(1-\cos\alpha\right)\ell\right]+\frac{\pi^2}{\ell} \sin^2\left(\frac{\alpha}{2}\right)\cos\left[\left(1+\cos\alpha\right)\ell\right]\;.
\end{eqnarray}
The analytical formula in Eq.~(\ref{eq:analytical}) is very efficient to compute the background potential at a long distance.
In Fig.~\ref{fig:compare} we compare the numerical results computed from Eq.~(\ref{eq:I_full}) with the analytical results from Eq.~(\ref{eq:analytical}). It can been seen that they match extremely well for $\ell \gg 1$.
Recalling $\ell=r E_\nu$   the background potential at a long distance is given by
\begin{eqnarray}
V_{\rm bkg}\left(r\gg E_\nu^{-1},\alpha\right)&=&-\frac{g_V^1 g_V^2}{\pi}G_F^2  \Phi_0 E_\nu \frac{1}{r}\left\{\cos^2\left(\frac{\alpha}{2}\right)\cos\left[\left(1-\cos\alpha\right)E_\nu r\right]\right.\nonumber\\
&&\left.
+\sin^2\left(\frac{\alpha}{2}\right)\cos\left[\left(1+\cos\alpha\right)E_\nu r\right]\right\}\;.
\label{eq:V-large-r}
\end{eqnarray}
We further consider the small $\alpha$ limit ($\alpha\ll 1$ while $E_\nu r \alpha^2$ can be arbitrarily large) and find
%. Yet, we keep the product $\ell \alpha$ arbitrary.
% We find
\begin{equation}
	\label{eq:1/r potential}
V_{\rm bkg}\left(r\gg E_\nu^{-1},\alpha \ll 1\right)=-\frac{g_V^1 g_V^2}{\pi}G_F^2  \times \Phi_0 E_\nu \times \frac{1}{r}\times \cos\left(\frac{\alpha^2 E_\nu r}{2}\right).
\end{equation} 

A few remarks are in order:
\begin{itemize}
\item 
The first term depends on the couplings of the fermions to the neutrinos.
\item
The second term is the energy density of the background neutrinos.
\item
The third term is the leading $r$ dependence. We learn that we have a $1/r$ potential.
\item
The last term encodes the angular dependency. We discuss it in more detail below.
\item
To leading order, this effect has no mass dependence. This is because the mass of the neutrino is negligible compared to the energies of the background neutrinos.
\end{itemize}

We next move to discuss the forces between macroscopic objects. In that case, we need to integrate over the energy of the background neutrinos as well as over the distribution of the masses. This integration can result in a smearing of the force, leading to the oscillatory behavior averaging out as we span the size of the macroscopic objects.

In order to get an effective $1/r$ potential, the smearing should not be very strong. 
The $\alpha$-suppressed oscillation mode starts to rapidly
oscillate when $\alpha^{2}\,\Delta (E_{\nu}\,r)\sim\pi$, where $\Delta (E_\nu r)$ is the spread of the energy $E_\nu$ and the location of the test masses. So the $1/r$ dependence
approximately holds if
\begin{equation}
\alpha^{2}\lesssim\frac{\pi}{\Delta( E_{\nu}\,r)}\thinspace.\label{eq:alpha-crit}
\end{equation}

\subsection{Discussion}

The neutrino-force effect is most significant when the background has a direction. There are several significant differences when comparing it to the vacuum case:
\begin{enumerate}
\item
\textbf{$r$ dependence.} While in vacuum the force scales as $1/r^5$, the leading term for a directional background scales as $1/r$. This implies that, at large distances, the background effects always overcome the vacuum contribution. Moreover, it implies that this force scales like gravity and the Coulomb force.
\item
\textbf{Oscillation.} The force exhibits oscillatory behavior. The oscillation length depends on the energy of the background neutrinos and the angle spanned by the background's direction and the direction of the induced force. Only at $\alpha=0$ there is no oscillation. 
\end{enumerate}

We provide some intuition for these two effects below. (Some of the discussions below
are based on ref.~\cite{VanTilburg:2024tst}).
The point is that, in the presence of background neutrinos, one of the virtual neutrinos in the loop is effectively replaced by a real neutrino, as imposed by the delta function $\delta(k^2-m^2)$ in the background propagator. Then, roughly speaking, the potential is related to the forward scattering amplitude of the real neutrinos between the two objects that are subject to the force. Usually, in the absence of a background, the mass suppression is a result of the ``off-shellness'' from the momentum transfer $q^2$. But in the presence of the high energy directional background, the departure from ``off-shellness'' is not so straightforward. The situation in vacuum is Lorentz invariant so the departure from $q^2$ is simply $m^2$. In the presence of the directional background, Lorentz-noninvariant quantities can be present in the propagator, which is what happens in this case.

Thus, in the vacuum case for a one-particle exchange potential, the potential is the Fourier transform of $(\mathbf{q}^2+m^2)^{-1} $, yielding $e^{-mr}/r$. In the background the propagator $\Pi (|\mathbf{q}|^2) $  is given by (as in Eq.~\eqref{eq:1011}):
\beq \Pi (|\mathbf{q}|^2) \sim  {1 \over \mathbf{q}^2 - 4E_{\nu}^2 \cos^2\theta_{\mathbf{k}_0 ,\mathbf{q}}} = {1 \over \mathbf{q}^2 + (2 i E_{\nu} \cos\theta_{\mathbf{k}_0 ,\mathbf{q}})^2 }, \eeq
where $\theta_{\mathbf{k}_0 ,\mathbf{q}}$ is the angle between vectors $\mathbf{k}_0$ and $\mathbf{q}$ ($\xi \equiv \cos\theta_{\mathbf{k}_0 ,\mathbf{q}}$ in Eq.~\eqref{eq:defxi}). The propagator has no leading order dependence on $m_{\nu}$ since $E_{\nu} \gg m_{\nu}$. Note that the ``off-shellness", which is real (i.e, $m^2$) in the vacuum case, is now imaginary in the presence of the background. 

We naively therefore obtain a Fourier transform,
\beq V(r) \sim { e^{- 2 i E_{\nu} r f(\alpha)} \over r}  \sim {1 \over r} \cos (2 E_{\nu} r f(\alpha) ), \label{eq:naiveFT} \eeq
where $f(\alpha) $ is some function of the angle $\alpha$, which we cannot predict without performing the integral explicitly.  This rough form allows us to intuit the features of the potential:

\begin{enumerate}
\item 
The $1/r$ dependence is the geometrical factor for an exchange of a massless intermediate particle. The background neutrinos practically make the potential from a two body exchange into a one body exchange, as evident from Eq.~\eqref{eq:1011}. One of the neutrinos is not virtual.
\item
The oscillation behavior arises from the fact that the background neutrinos modify the propagator to carry an imaginary ``mass term". This makes the exchanged neutrino ``real" as opposed to virtual, giving an oscillatory behavior. Another way this can be understood is as an interference effect between two amplitudes. One amplitude is the incoming background wave and the other one that scatters off one of the two interacting objects. At large $r$, for $\alpha =0$, the interference is pure constructive and the potential behaves as $1/r$, corresponding to $f(0)=0$ in Eq.~\eqref{eq:naiveFT} above. While at $\alpha =\pi/2$, there is destructive interference and oscillatory behavior persists.

\end{enumerate}

%%%%%%%%%%%%%%%%%%%%%%%

\section{Experimental sensitivities and detection of neutrino forces}
\label{sec:exp}
\subsection{Current status of the experiments}
There have been decades of experimental efforts to search for new
long-range forces (also referred to as the fifth force) -- see Refs.~\cite{Wagner:2012ui,Adelberger:2009zz,fifth-force}
for reviews. Searches that typically employ torsion balance devices
are closely related to precision tests of gravity, more specifically,
to tests of the gravitational inverse-square law (ISL)~\cite{Hoskins:1985tn, Lee:2020zjt, Tan:2020vpf}
and tests of the weak equivalence principle (WEP)~\cite{Schlamminger:2007ht,Smith:1999cr}.
We summarize the experimental sensitivities in Table~\ref{tab:exp_list}
and compare them with our theoretical expectations of neutrino forces including background corrections
in Fig.~\ref{fig:Exp}. The details are explained in what follows. 

Experiments testing the WEP look for possible  differences between the
accelerations of different test bodies in the same gravitational field.
For example, the gravitational acceleration on the Earth, $a_{\oplus}\approx9.8\ {\rm m}/{\rm s}^{2}$,
should be universal for all test bodies at the same location, independent
of the material of the test body. In the presence of a new long-range
force whose  couplings to electrons and nucleons are disproportional
to their masses, the actual observed acceleration may violate the
universality.  

%%%%%%%%%%%%%%%%%%%%%%%%%%%%%%% Table 4 %%%%%%%%%%%%%%%%%%%%%%%%%%
\begin{table*}
	\centering
	
	\begin{tabular}{ccccc}
		\toprule 
		exp & $\delta V/V_{{\rm gravity}}$ & $\langle r\rangle$ & Refs & \tabularnewline
		\midrule 
		Washington2007 & $3.2\times10^{-16}$ & $\sim6400$ km & \cite{Schlamminger:2007ht} & \tabularnewline
		Washington1999 & $3.0\times10^{-9}$ & $\sim0.3$ m & \cite{Smith:1999cr} & \tabularnewline
		Irvine1985 & $0.7\times10^{-4}$ & $2-5$ cm & \cite{Hoskins:1985tn} & \tabularnewline
		Irvine1985 & $2.7\times10^{-4}$ & $5-105$ cm & \cite{Hoskins:1985tn} & \tabularnewline
		Wuhan2012 & $10^{-3}$ & $\sim2$ mm & \cite{Yang:2012zzb} & \tabularnewline
		Wuhan2020 & $3\times10^{-2}$ & $\sim0.1$ mm & \cite{Tan:2020vpf} & \tabularnewline
		Washington2020 & $\sim1$ & $52$ $\mu$m & \cite{Lee:2020zjt} & \tabularnewline
		Future levitated optomechanics & $\sim10^{4}$ & 1 $\mu$m & \cite{Moore:2020awi} & \tabularnewline
		\bottomrule
	\end{tabular}
	
	\caption{\label{tab:exp_list} Sensitivities of long-range force search experiments. }
\end{table*}
%%%%%%%%%%%%%%%%%%%%%%%%%%%%%%%%%%%%%%%%%%%%%%%%%%%%%%%%%%%%%%%%%

Using Be and Ti as test masses and measuring the difference between
their gravitational accelerations, the Washington experiment group
reported the following result in 2007~\cite{Schlamminger:2007ht}:
\begin{equation}
a_{{\rm Be}}-a_{{\rm Ti}}=(0.6\pm3.1)\times10^{-15}\ {\rm m}/{\rm s}^{2}\thinspace\ \ ({\rm Earth\ attractor})\thinspace.\label{eq:a-earth}
\end{equation}
Here, the Earth serves as the gravitational attractor. The average
distance between particles in the test body and in the attractor in
this case is roughly the radius of the Earth, $\langle r\rangle\sim6400\ {\rm km}$.
Dividing the experimental uncertainty in Eq.~\eqref{eq:a-earth}
by $a_{\oplus}\approx9.8\ {\rm m}/{\rm s}^{2}$, we obtain $\delta V/V_{{\rm gravity}}=3.2\times10^{-16}$
where $V_{{\rm gravity}}$ is the gravitational potential and $\delta V$
denotes potential variations due to new forces. This experimental
setup is referred to as Washington2007 in Table~\ref{tab:exp_list}. 

Instead of making use of the Earth's gravity, one can also employ
laboratory attractors. An earlier experiment conducted by the same
group using a 3-ton $^{238}$U attractor and test bodies of Cu and
Pb reported~\cite{Smith:1999cr}:
\begin{equation}
a_{{\rm Cu}}-a_{{\rm Pb}}=(1.0\pm2.8)\times10^{-15}\ {\rm m}/{\rm s}^{2}\ \ \ \ (\text{3-ton\ \ensuremath{{}^{238}}U attractor})\thinspace.\label{eq:a-U}
\end{equation}
Note that the uncertainty is close to the one in Eq.~\eqref{eq:a-earth}
but the result should be compared with the gravitational acceleration
caused by the $^{238}$U attractor, which is $9.2\times10^{-7}\ {\rm m}/{\rm s}^{2}$.
The $^{238}$U attractor has an annular shape with inner and outer
radii of 10.2 cm and 44.6 cm while the torsion balance is located
in its center.  Hence the average distance between particles in the
test body and in the attractor in this case is roughly $\langle r\rangle\sim0.3\ {\rm m}$.
This experimental setup is referred to as Washington1999 in Table~\ref{tab:exp_list}.

Experiments testing ISL measures the variation of the gravitational
attraction between two test bodies when their distance varies. The
Irvine experiment conducted in the 1980s was already able to probe
ISL over a distance range from 2 cm to 105 cm at the precision of
$10^{-4}$\cite{Hoskins:1985tn}, ruling out a previously claimed
deviation of ISL by $(0.37\pm0.07)\%$ in the 4.5 to 30 cm range~\cite{long1976experimental}.
In recent years, the precision of ISL testing experiments in the centimeter
to meter range has not been improved significantly. The main progress
that has been made so far is the successful measurement of gravitational
forces at much smaller distance scales~\cite{Lee:2020zjt,Tan:2020vpf}.
So far, the smallest distance scale at which gravity has been probed
in laboratory is $52\ {\rm \mu m}$~\cite{Lee:2020zjt}.   Above
this scale, gravitational forces have been measured to certain precision
(see results of Wuhan2012~\cite{Yang:2012zzb}, Wuhan2020~\cite{Tan:2020vpf},
and 
Washington2020~\cite{Lee:2020zjt} in Table~\ref{tab:exp_list})
and the measurements are fully consistent with ISL. 

%%%%%%%%%%%%%%%%%%%%%%%%%%%%%
\subsection{Detection of neutrino forces}

When applying the above experimental sensitivities to neutrino forces,
one should note that $\delta V$ caused by reactor and solar neutrinos
are both direction-dependent. For solar neutrinos, the angle $\alpha$ 
varies with a period of 24 hours
due to Earth's rotation. For reactor neutrinos, the angle $\alpha$ varies
in experiments with moving attractors, as is the case of the Washington1999
experiment~\cite{Smith:1999cr}. Since the reactor neutrino flux
is only intense within a short distance from a reactor, the Washington2007
experiment does not provide strong probing power to the reactor neutrino force. 

In order to compare the deviation of ISL gravitational potential from the background potentials to the experimental sensitivities, we need to compute $V_{\rm bkg}$ between two objects numerically and compare it to the gravity. As a bench mark point, we fix $\alpha=0$.

This assumption is not valid in all of the examples that we study below. All of the current experiments are done between extended objects and the averaging over their shape is important, making the use of the $\alpha=0$ result unjustified. Yet, we do use the $\alpha=0$ as the most optimistic scenario just to get an idea how far the effects are from current sensitivities.

Since in the cases we are considering, the vacuum potential is negligible the neutrino force between two particles in the directional neutrino background is simply given by
\begin{equation}
V_{\nu{\rm -force}}(r)=-\frac{g^1_V g^2_V}{\pi}G_{F}^{2}\Phi E_{\nu}\frac{1}{r}\qquad \left(r\gg E_\nu^{-1},~\alpha=0\right)\;,
\label{eq:V_for_plot}
\end{equation}
which is proportional to $1/r$, same as the gravitational potential. Notice that the typical energy of reactor and solar neutrino flux is $E_\nu\sim {\rm MeV}\sim \left(10^{-11}{\rm cm}\right)^{-1}$, while the average distance between two particles in the test body and in the attractor is larger than ${\rm \mu m}$ (cf. Table~\ref{tab:exp_list}). Hence, we only need to consider the long-range behavior of the background potential, namely, $r\gg E_\nu^{-1}$. We use Eq.~(\ref{eq:V_for_plot}) below to compute the background potentials.

%%%%%%%%%%%%%%%%%%%%%%%%%%%  Fig.5 %%%%%%%%%%%%%%%%%%%%%%%%%%%%%%%%%%%%%
\begin{figure}
\centering	
\includegraphics[width=0.85\textwidth]{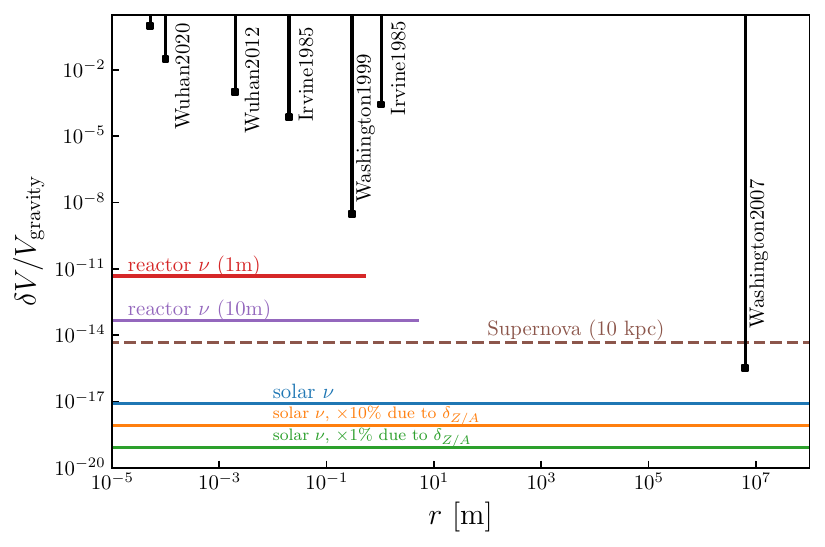}\caption{Neutrino forces in comparison with experimental sensitivities. Here all neutrino sources are assumed to be ideally point-like and the angular spread is assumed to be sufficiently small to meet Eq.~\eqref{eq:alpha-crit}. In reality, a sizable angular spread needs to be taken into account thus the above should be considered as an upper bound of the effect.  \label{fig:Exp}}
\end{figure}
%%%%%%%%%%%%%%%%%%%%%%%%%%%%%%%%%%%%%%%%%%%%%%%%%%%%%%%%%%%%%%%%%%%%%%%%

In Fig.~\ref{fig:Exp}, we plot the reactor neutrino force curves calculated from Eq.~(\ref{eq:V_for_plot})
using the standard reactor neutrino flux at 1 meter and
10 meters from the reactor core. For a reactor with 2.9 GW thermal power, the neutrino flux is $\Phi=5\times 10^{13}\ {\rm cm}^{-2}{\rm s}^{-1}$ at 10 meters away~\cite{Kopeikin:2012zz}. We take $E_\nu$ in Eq.~(\ref{eq:V_for_plot}) to be 2 MeV when computing the background potential from reactor neutrinos.  
The curves stop at $r=0.5$ m and $r=5$ m because experiments with much larger $r$ (such as Washington2007) are impossible to have test bodies and attractors all fitted in the limited space within 1 or 10 meters from the reactor.  

For solar neutrinos, this is not a concern. So far, all experiments have $r$ much smaller than the distance to the Sun. However,
one should note that the angle $\alpha$ varies with a period of 24 hours while a large number of noises are also 24-hour periodic. Hence the $\alpha$ dependence could be easily submerged in such noises. Nevertheless, we plot the solar neutrino line in Fig.~\ref{fig:Exp} assuming that it could be resolved among various noises in future experiments. 

The solar neutrino line in Fig.~\ref{fig:Exp} is calculated from Eq.~(\ref{eq:V_for_plot}) by considering $pp$ neutrinos with the flux $\Phi=5.99\times 10^{10} {\rm cm}^{-2} {\rm s}^{-1}$ and the highest energy $E_{\rm max}=0.42 {\rm \,MeV}$~\cite{Vinyoles:2016djt}. In the computation we take $E_\nu=0.3 {\rm\, MeV}$ since the $pp$ neutrino spectrum is not monochromatic. We have also calculated the background potential of the $^{7}$Be solar neutrinos whose flux is $\Phi=4.84\times 10^{9} {\rm \,cm}^{-2} {\rm s}^{-1}$ with two monochromatic energies being $E_\nu=0.862 {\rm \,MeV}$ and $E_\nu=0.384 \,{\rm MeV}$~\cite{Vinyoles:2016djt}. But the result is the same order of magnitude as that of $pp$ neutrinos.

It might be more feasible to make use of the material dependence feature
of neutrino forces. Since the effective neutrino-proton vector coupling
is suppressed by a factor of $1-4\sin^2\theta_{W}\approx0.05$ with
respect to the effective neutrino-neutron vector coupling, we can
assume that neutrino forces mainly depend on the neutron number $N=A-Z$
($A$: atomic mass number, $Z$: proton number) of the material used
in test bodies. The contribution of electrons is more complicated
since the charged-current interaction may or may not contribute (if
not, the $1-4\sin^2\theta_{W}$ suppression also applies to electrons),
depending on the neutrino flavor. For simplicity, here we neglect the electron contribution (see Appendix~\ref{app:ZA} for a more strict treatment).
Therefore, for neutrino forces on different materials, the difference is roughly
\begin{equation}
\frac{\delta V_{\nu{\rm -force}}}{V_{\nu{\rm -force}}}\sim\delta_{Z/A},\ \delta_{Z/A}=\begin{cases}
1.6\% & \text{for Be vs Ti }\\
4.9\% & \text{for Cu vs Pb }\\
8.2\% & \text{for Al vs Pt }
\end{cases}.\label{eq:x}
\end{equation}
Here $Z/A$ is approximately $1/2$ for most nuclei, and $\ \delta_{Z/A}$
denotes its variation for different materials. Taking Be vs Ti for
example, since Ti (Be) has 22 (4) protons and 26 (5) neutrons, the
difference is $22/48-4/9=1.4\%$. More accurate calculations using
$A=47.87$ ($9.012$) gives $1.6\%$. In principle, $\delta_{Z/A}$
could be enhanced to as large as $50\%$ if Hydrogen ($Z=A=1$) is
used in combination with other $Z/A\approx1/2$ material, though technically
it is difficult to make test bodies of Hydrogen. In Fig.~\ref{fig:Exp},
below the solar neutrino line,
we plot two lines by multiplying it with $\delta_{Z/A}=10\%$ and
$1\%$. If the direction-dependent signal of $V_{\nu{\rm -force}}$
are submerged in various 24-hour noises, the material dependence of
$V_{\nu{\rm -force}}$, which is a factor of $\delta_{Z/A}$ weaker
but more robust against noises, could be exploited to probe neutrino
forces.

In addition to the aforementioned dependence on directions and materials, the difference between reactor-on and -off measurements could also be used to probe neutrino forces.

For supernova neutrinos, we plot a dashed line in Fig.~\ref{fig:Exp} to present the magnitude.  We assume that the supernova neutrino flux is $10^{12}\ {\rm cm}^{-2}{\rm s}^{-1}$, corresponding to a 10 kpc core-collapse supernova neutrino burst~\cite{Kachelriess:2004ds}. The neutrino mean energy is about $10$ MeV. 
Here we use a dashed line to remind the readers that such a neutrino burst lasts only for a short period of a few seconds, which might be too short for torsion balance experiments to reach the desired sensitivity (e.g.~the torsional oscillation period of Washington 2007 is 798 s~\cite{Schlamminger:2007ht}).  A dedicated analysis on such experiments taking the short duration into consideration might lead to a much weaker sensitivity, but this is beyond the scope of our work.

At last, we give some brief remarks on the background effects from atmospheric and accelerator neutrinos.
The flux of atmospheric neutrinos is much smaller than those  of the reactor and solar neutrinos~\cite{Raffelt2020}. As a result, the corresponding background potential is weaker than that of reactor neutrinos by 12 orders of magnitude. In addition, the flux from long-baseline accelerator neutrino experiments like DUNE~\cite{DUNE:2020lwj} is also weaker than that of reactor neutrinos. The accelerator neutrino background potential at the near-detector location of DUNE is about 7 orders of magnitude smaller than that of reactor neutrinos. Therefore, the background potentials from both atmospheric and accelerator neutrinos are out of the reach with current experimental sensitivities.

\section{Conclusions}
\label{sec:summary}
In this paper, we computed the background corrections to neutrino forces in a thermal or non-thermal neutrino background. We found that the presence of the background can significantly increase the strength of neutrino forces.

For the isotropic C$\nu$B in Eq.~(\ref{eq:FDdistribution2}), we have derived general formulae of the background potential for both Dirac [Eq.~(\ref{eq:VTFDdistribution})] and Majorana  [Eq.~(\ref{eq:VTFDdistributionMajorana})] neutrinos that are valid for arbitrary neutrino masses and distances. The main feature of the  potential in the presence of the  C$\nu$B is that, at large distances ($r\gg m_\nu^{-1}$), it is not exponentially suppressed, as opposed to the potential in vacuum. Therefore, when the distance between two particles exceeds the inverse mass of neutrinos, the neutrino force between them is dominated by the background contribution. However, since the number density of the cosmic neutrinos is very small today, the thermal effects of the C$\nu$B
on the neutrino force are still far from the available experimental sensitivities.

We then computed the neutrino force in a directional background. 
%a much more dense background of solar and reactor neutrinos. 
We parametrized the non-thermal and anisotropic background as monochromatic distribution function with a specific direction, $\alpha$. The general direction-dependent background potential is given by Eqs.~(\ref{eq:anisotropicpotential}) and (\ref{eq:I_full}). At $r\gg E_\nu^{-1}$ with 
%$E_\nu^{}\sim {\rm MeV}$ 
$E_\nu^{}$ being the typical energy of the neutrino flux, the background potential in the small $\alpha$ limit is proportional to $1/r$, which falls much slower  than the $1/r^5$ potential in vacuum and in isotropic backgrounds. In particular, there is a potential significant enhancement of the vacuum force in the presence of directional energetic dense neutrino backgrounds. 
%For example, for two objects separated by $r=1$ meter, the potential in the reactor neutrino background at
%1 meter away from the reactor core
%calculated using Eq.~(\ref{eq:V_for_plot}) is \emph{26 orders of magnitude larger} than the vacuum potential in Eq.~(\ref{eq:vacuumpotential}). 

%However, we would like to remind the readers that this enhancement is valid only when the non-smearing condition of Eq.~(\ref{eq:alpha-crit}) is satisfied, otherwise the leading $1/r$ potential would be smeared by the finite energy and angular spread. Although it is difficult for existing experiments to satisfy Eq.~(\ref{eq:alpha-crit}), theoretically, a strong enhancement is possible. 

%Since the neutrino force in an energetic and dense neutrino background is significantly enhanced, 
We then turned to discuss the possibility of probing the neutrino force using torsion balance experiments that aim to precisely test the gravitational inverse-square law and the weak equivalence principle. 
Assuming the small $\alpha$ limit, the comparison of the neutrino force in reactor and solar neutrino backgrounds with experimental sensitivities is summarized in Fig.~\ref{fig:Exp}. 
The figure shows that, if  Eq.~(\ref{eq:alpha-crit}) could be satisfied, the current experiments would be 2 or 3 orders of magnitude far from detecting neutrino forces in the reactor or solar neutrino background.
With current technology, however, the condition in Eq.~(\ref{eq:alpha-crit})  is not satisfied and
%However, we would like to remind the readers that it is difficult for existing experiments to satisfy Eq.~(\ref{eq:alpha-crit}) and 
the  energy and angular spread smear out the leading $1/r$ potential.  
While it is not clear to us how complicated and practical it is to design an experiment that can exploit the enhancement we discuss,
the point to emphasize is that strong enhancement is present. %theoretically possible and this is a substantial improvement compared with previous studies on possible approaches to the detection of neutrino forces.

We conclude that the neutrino force in the solar or reactor neutrino background is much more experimentally accessible than the one in vacuum.  Dedicated experimental efforts are called for to check if these enhancement factors can be exploit in order to detect the elusive neutrino force.

\vspace{0.2cm}

\noindent\textbf{Note added.} After we updated our paper on arXiv to Version 2, Ref.~\cite{Blas:2022ovz} appeared on arXiv. The authors of that preprint commented that the finite size of the wave packets would destroy the leading $1/r$ potential in directional neutrino backgrounds that we found. However, the content of \cite{Blas:2022ovz} was referring to Version 1 of our paper, while in Version 2 we have already addressed the smearing effect. To address the effect of the smearing, 
Ref.~\cite{Blas:2022ovz} took a different approach than ours. They included the energy spread in the wave packets first and then took a monochromatic directional flux and fixed $\alpha=0$, while we consider the smearing effect by varying $E_\nu$ and $\alpha$ of the flux. 
%The two approaches lead to the same conclusion that a finite energy spread combined with a finite angular spread would smear the $1/r$ potential in Eq.~\eqref{eq:1/r potential}.
While the details of our analyses are not identical, the results of the current version of our work are in agreement with the results of Ref.~\cite{Blas:2022ovz}.
Yet, our conclusions have a different tone. While we emphasize the fact that there is indeed a strong enhancement when Eq.~(\ref{eq:alpha-crit}) is satisfied, Ref.~\cite{Blas:2022ovz} is worried  about the feasibility of designing experiments that can use it.

\section*{Acknowledgements}
We thank   Ken Van Tilburg for discussions and for sharing with us
results from his soon to be completed work~\cite{VanTilburg:2024tst}.
We also thank  Ephraim Fischbach,
Concha Gonzales-Garcia, Dennis Krause, and Enrico Nardi, for remarks and discussions. 
The work of YG is supported in part by the NSF grant PHY1316222.  W.T. is supported by the NSF Grant No. PHY-2013052. The work of BY is supported by the National Natural Science Foundation of China under grant No. 11835013. X.J.X is supported in part by the National Natural Science Foundation of China under grant No. 12141501.

\begin{appendix}

\section{The background effect on fermion propagators \label{sec:background-effect}}

The neutrino propagator in a background with finite neutrino number
density in Eq.~\eqref{eq:S} can be found in various references including
books and reviews on finite temperature field theory~\cite{Landsman:1986uw,Notzold:1987ik,Quiros:1999jp,Kapusta:2006pm,Laine:2016hma}. In this appendix,
we provide a simple and pedagogical re-derivation of the formula {\it
	without using finite temperature field theory}, aiming at providing
a physical interpretation of the background effect. 

Let us start with the propagator of a generic fermion in vacuum, which
is defined as 
\begin{equation}
S_{F}(x-y)\equiv\langle0|T\psi(x)\overline{\psi(y)}|0\rangle\thinspace,\label{eq:x-3}
\end{equation}
where $T$ indicates that it is a time-ordered product. Using
\begin{equation}
\psi=\int\frac{d^{3}\mathbf{p}}{(2\pi)^{3}}\frac{1}{\sqrt{2E_{\mathbf{p}}}}\sum_{s}\left[a_{\mathbf{p}}^{s}u^{s}(p)e^{-ip\cdot x}+b_{\mathbf{p}}^{s\dagger}v^{s}(p)e^{ip\cdot x}\right],\label{eq:x-4}
\end{equation}
where we follow the standard notation of Ref.~\cite{Peskin},
%Peskin\&Schroeder's book~\cite{Peskin}
and assuming $x^{0}>y^{0}$ so that $T$ can be removed, we obtain
\begin{align}
S_{F} & \propto\int\frac{d^{3}\mathbf{p}}{(2\pi)^{3}}\int\frac{d^{3}\mathbf{k}}{(2\pi)^{3}}\frac{1}{\sqrt{2E_{\mathbf{p}}}}\frac{1}{\sqrt{2E_{\mathbf{k}}}}e^{-ip\cdot x+ik\cdot y}\langle0|a_{\mathbf{p}}a_{\mathbf{k}}^{\dagger}|0\rangle+\cdots,\label{eq:x-2}
\end{align}
where for brevity we have neglected $u^{s}$, $v^{s}$, and the script
$s$ (they only affect the structure of Dirac spinors). The ``$\cdots$''
denote terms proportional to $\langle0|a_{\mathbf{p}}b_{\mathbf{k}}|0\rangle,$
$\langle0|b_{\mathbf{p}}^{\dagger}a_{\mathbf{k}}^{\dagger}|0\rangle$,
or $\langle0|b_{\mathbf{p}}^{\dagger}b_{\mathbf{k}}|0\rangle$, all
being zero. Since $\langle0|a_{\mathbf{p}}a_{\mathbf{k}}^{\dagger}|0\rangle=(2\pi)^{3}\delta^{3}(\mathbf{p}-\mathbf{k})$,
Eq.~\eqref{eq:x-2} gives 
\begin{align}
S_{F}\propto & \int\frac{d^{3}\mathbf{p}}{(2\pi)^{3}}\frac{1}{2E_{\mathbf{p}}}e^{-ip\cdot(x-y)}=\int\frac{d^{4}p}{(2\pi)^{4}}\frac{i}{p^{2}-m^{2}+i\epsilon}e^{-ip\cdot(x-y)}\thinspace.\label{eq:x-5}
\end{align}
The last step is simply the reverse process of computing the contour
integral of $p^{0}$, with the underlying assumption that $x^{0}>y^{0}$.
For $x^{0}<y^{0}$, the time ordering guarantees the same result. 

Now we shall replace $|0\rangle$ with a background state. Let us first consider a single-particle state which contains a particle
with an almost certain position and an almost certain momentum. The
two cannot be simultaneously fixed at exact values due to the uncertainty
principle, but one can nevertheless introduce a wave package function
$w(\mathbf{p})$ so that both $w(\mathbf{p})$ and its Fourier transform
$\int w(\mathbf{p})e^{i\mathbf{p}\cdot\mathbf{x}}d^{3}\mathbf{x}$
are limited in a small region of their respective space---for further
elucidation, see e.g.~Appendix A of Ref.~\cite{Smirnov:2022sfo}. The single
particle state is then defined as 
\begin{equation}
|w\rangle=\int\frac{d^{3}\mathbf{p}}{(2\pi)^{3}}w(\mathbf{p})a_{\mathbf{p}}^{\dagger}|0\rangle\thinspace,\ \langle w|w\rangle=\int\frac{d^{3}\mathbf{p}}{(2\pi)^{3}}|w(\mathbf{p})|^{2}\equiv1\thinspace,\label{eq:x-6}
\end{equation}
where the last step is defined as the normalization condition of
$w(\mathbf{p})$. 

Replacing $|0\rangle\to|w\rangle$ in Eq.~\eqref{eq:x-2}, we obtain
\begin{equation}
S_{F}\propto\int_{\mathbf{p}\mathbf{k}}\langle w|a_{\mathbf{p}}a_{\mathbf{k}}^{\dagger}|w\rangle=\int_{\mathbf{p}\mathbf{k}}\left\langle w\left|\left((2\pi)^{3}\delta^{3}(\mathbf{p}-\mathbf{k})-a_{\mathbf{k}}^{\dagger}a_{\mathbf{p}}\right)\right|w\right\rangle\thinspace,\label{eq:x-7}
\end{equation}
where $\int_{\mathbf{p}\mathbf{k}}$ standards for $\int\frac{d^{3}\mathbf{p}}{(2\pi)^{3}}\int\frac{d^{3}\mathbf{k}}{(2\pi)^{3}}\frac{1}{\sqrt{2E_{\mathbf{p}}}}\frac{1}{\sqrt{2E_{\mathbf{k}}}}e^{-ip\cdot x+ik\cdot y}$.
Since $\langle w|w\rangle=1$, the first term leads to the same result
as the vacuum case and the second term represents the background effect.
We denote the contribution of the latter by $S_{F}^{{\rm bkg}}$:
\begin{equation}
S_{F}^{{\rm bkg}}\propto-\int_{\mathbf{p}\mathbf{k}}\langle w|a_{\mathbf{k}}^{\dagger}a_{\mathbf{p}}|w\rangle=-\int_{\mathbf{p}\mathbf{k}}w^{*}(\mathbf{k})w(\mathbf{p})\thinspace,\label{eq:x-8}
\end{equation}
where we have used $a_{\mathbf{p}}|w\rangle=w(\mathbf{p})|0\rangle$.
Note that $w(\mathbf{p})$ has been defined in such a way that the
particle's position and momentum are nearly fixed at certain values
(say $\mathbf{x}_{0}$ and $\mathbf{p}_{0}$). One can perform spatial
translation of the wave package $w(\mathbf{p})\to w_{\Delta\mathbf{x}}(\mathbf{p})\equiv e^{i\mathbf{p}\cdot\Delta\mathbf{x}}w(\mathbf{p})$
so that its position is changed to $\mathbf{x}_{0}+\Delta\mathbf{x}$
while the momentum is unchanged. Now, if we randomly choose $\Delta\mathbf{x}$
with a uniform probability distribution in a large volume $V$ (much
larger than the distribution of each wave package), the position of
the particle would be evenly smeared in $V$. For $w^{*}(\mathbf{k})w(\mathbf{p})$
in Eq.~\eqref{eq:x-8}, the smearing leads to
\begin{align}
w^{*}(\mathbf{k})w(\mathbf{p})\xrightarrow{\text{smearing}} & \frac{1}{V}\int w_{\Delta\mathbf{x}}^{*}(\mathbf{k})w_{\Delta\mathbf{x}}(\mathbf{p})d^{3}\Delta\mathbf{x}\nonumber \\
& =\frac{1}{V}\int w^{*}(\mathbf{k})w(\mathbf{p})e^{i(\mathbf{p}-\mathbf{k})\cdot\Delta\mathbf{x}}d^{3}\Delta\mathbf{x}\nonumber \\
& =\frac{(2\pi)^{3}\delta^{3}(\mathbf{p}-\mathbf{k})}{V}|w(\mathbf{p})|^{2}\nonumber \\
& =(2\pi)^{3}\delta^{3}(\mathbf{p}-\mathbf{k})n_{+}(\mathbf{p})\thinspace,\label{eq:x-9}
\end{align}
where in the last step we have identified $|w(\mathbf{p})|^{2}/V$
as $n_{+}(\mathbf{p})$ because $\int\frac{d^{3}\mathbf{p}}{(2\pi)^{3}}|w(\mathbf{p})|^{2}=1$
and the number density after smearing is $\int\frac{d^{3}\mathbf{p}}{(2\pi)^{3}}n_{+}(\mathbf{p})=1/V$. 

Substituting Eq.~\eqref{eq:x-9} into Eq.~\eqref{eq:x-8}, we obtain
\begin{align}
-\int_{\mathbf{p}\mathbf{k}}w^{*}(\mathbf{k})w(\mathbf{p})\xrightarrow{\text{smearing}} & -\int\frac{d^{3}\mathbf{p}}{(2\pi)^{3}}\frac{e^{-ip\cdot(x-y)}}{2E_{\mathbf{p}}}n_{+}(\mathbf{p})\nonumber \\
& =-\int\frac{d^{4}p}{(2\pi)^{4}}e^{-ip\cdot(x-y)}(2\pi)\delta\left(p^{2}-m^{2}\right)\Theta\left(p^{0}\right)n_{+}(\mathbf{p})\thinspace.\label{eq:x-10}
\end{align}
Combining Eq.~\eqref{eq:x-10} with the vacuum part in Eq.~\eqref{eq:x-5},
we obtain
\begin{equation}
S_{F}\propto\int\frac{d^{4}p}{(2\pi)^{4}}e^{-ip\cdot(x-y)}\left\{ \frac{i}{p^{2}-m^{2}+i\epsilon}-(2\pi)\delta\left(p^{2}-m^{2}\right)\Theta\left(p^{0}\right)n_{+}(\mathbf{p})\right\} \thinspace.\label{eq:x-11}
\end{equation}
For an anti-particle background, the above calculation is similar
except that some minus signs are flipped.  In the presence of both
particles and anti-particles in the background, we obtain 
\begin{align*}
S_{F}(p) & =\left(\slashed{p}+m\right)\left\{ \frac{i}{p^{2}-m^{2}+i\epsilon}-(2\pi)\delta\left(p^{2}-m^{2}\right)\Theta\left(p^{0}\right)\left[n_{+}(\mathbf{p})-n_{-}(\mathbf{p})\right]\right\} \\
& =\left(\slashed{p}+m\right)\left\{ \frac{i}{p^{2}-m^{2}+i\epsilon}-(2\pi)\delta\left(p^{2}-m^{2}\right)\left[\Theta\left(p^{0}\right)n_{+}(\mathbf{p})+\Theta\left(-p^{0}\right)n_{-}(\mathbf{p})\right]\right\} ,
\end{align*}
where $S_{F}(p)$ is the propagator in the momentum space {[}i.e.~the
Fourier transform of $S_{F}(x-y)${]}, the prefactor $\left(\slashed{p}+m\right)$
can be inferred from the vacuum propagator. The result is the same
as the fermion propagator derived in finite temperature field theory. 

From the above calculation, one can see that the background effect
comes from the second term in Eq.~\eqref{eq:x-7}, proportional to
$\langle w|a_{\mathbf{k}}^{\dagger}a_{\mathbf{p}}|w\rangle$. Recall
that the annihilation operator $a_{\mathbf{p}}$ acting on $|w\rangle$
can be interpreted as reducing one particle in the background. Hence
$\langle w|a_{\mathbf{k}}^{\dagger}a_{\mathbf{p}}|w\rangle$ corresponds
to first absorbing a particle of momentum $\mathbf{p}$ from the background
($a_{\mathbf{p}}|w\rangle=w(\mathbf{p})|0\rangle$), and returning
a particle of momentum $\mathbf{k}$ back to the background. Smearing
the single particle state in Eq.~\eqref{eq:x-9} leads to $\delta^{3}(\mathbf{p}-\mathbf{k})$,
which guarantees that the particle being returned has the same momentum
as the one being absorbed.

Intuitively, the modified propagator in Eq.~(\ref{eq:S}) can be understood as the vacuum expectation value of two fermion fields with the vacuum state $|0\rangle$ replaced by the modified background state $|w\rangle$, which is the vacuum equipped with some on-shell background fermions. Then the Wick contraction can be carried out not only between the two internal fermion fields (leading to the vacuum propagator), but also among the internal fields and the background fermions (leading to the modified term). Therefore, the modified term is naturally proportional to the number density of background fermions, with the factor $2\pi \delta(p^2-m^2) \Theta(p^0)$ coming from cutting the propagator to put it on-shell (optical theorem). Notice that the above arguments should be valid in any background and do not require the distribution to be thermal.

\section{Integrals \label{sec:integral}}
In this appendix, we present the details about some integrals in calculating neutrino forces in the neutrino backgrounds.

\subsection{Derivation of the general background potential $V_{\rm bkg}(r)$ in Eq.~(\ref{eq:V_iso})}
We first show how to obtain the general expression of the background potential with an arbitrary distribution function. 

As has been stated above, when both neutrino propagators in Eq.~(\ref{eq:amplitude}) take the first part, it corresponds to the vacuum potential $V_0(r)$, which is independent of the background distribution functions. When both propagators take the second part, the result always vanishes because of the existence of two delta functions. Therefore, the background contribution comes from the cross terms, i.e., $S_\nu(k)$  takes the first (second) part and $S_{\nu}(k+q)$ takes the second (first) part:
\begin{eqnarray}
	\label{eq:AT1}
{\cal A}_{{\rm bkg}}(q) & = & -\pi G_{F}^{2}g^1_V g^2_V\int\frac{d^{4}k}{\left(2\pi\right)^{4}}\delta\left(k^{2}-m_\nu^{2}\right)\left[\Theta\left(k^{0}\right)n_{+}\left(\mathbf{k}\right)+\Theta\left(-k^{0}\right)n_{-}\left(\mathbf{k}\right)\right]\nonumber \\
&  & \times\left\{ \frac{{\rm Tr}\left[\gamma^{0}\left(1-\gamma_{5}\right)\left(\slashed{k}+m_\nu\right)\gamma^{0}\left(1-\gamma_{5}\right)\left(\slashed{k}+\slashed{q}+m_\nu\right)\right]}{\left(k+q\right)^{2}-m_\nu^{2}}\right.\nonumber \\
&  & \left.+\frac{{\rm Tr}\left[\gamma^{0}\left(1-\gamma_{5}\right)\left(\slashed{k}-\slashed{q}+m_\nu\right)\gamma^{0}\left(1-\gamma_{5}\right)\left(\slashed{k}+m_\nu\right)\right]}{\left(k-q\right)^{2}-m_\nu^{2}}\right\}\, , \;\nonumber \\
& = & -8\pi G_{F}^{2}g^1_V g^2_V\int\frac{d^{4}k}{\left(2\pi\right)^{4}}\delta\left(k^{2}-m_\nu^{2}\right)\left[\Theta\left(k^{0}\right)n_{+}\left(\mathbf{k}\right)+\Theta\left(-k^{0}\right)n_{-}\left(\mathbf{k}\right)\right]\nonumber \\
&  & \times\left[\frac{2k^{0}\left(k^{0}+q^{0}\right)-\left(k\cdot q+k^{2}\right)}{\left(k+q\right)^{2}-m_\nu^{2}}+\frac{2k^{0}\left(k^{0}-q^{0}\right)+\left(k\cdot q-k^{2}\right)}{\left(k-q\right)^{2}-m_\nu^{2}}\right]\;.
\end{eqnarray}
Taking advantage of the identity
\[
\delta\left(k^{2}-m_\nu^{2}\right)=\delta\left(\left(k^{0}\right)^{2}-E_{k}^{2}\right)=\frac{1}{2E_{k}}\left[\delta\left(k^{0}-E_{k}\right)+\delta\left(k^{0}+E_{k}\right)\right]\;,
\]
one can first integrate $k^{0}$ in Eq.~(\ref{eq:AT1}). In addition, the NR approximation requires $q\simeq\left(0,\mathbf{q}\right)$. Thus the integral in Eq.~(\ref{eq:AT1}) can be reduced to Eq.~(\ref{eq:Abkg})
\begin{eqnarray}
	\label{eq:AT2}
	{\cal A}_{{\rm bkg}}(\mathbf{q})=4G_{F}^{2}g^1_V g^2_V\int\frac{d^{3}\mathbf{k}}{\left(2\pi\right)^{3}}\frac{n_{+}\left(\mathbf{k}\right)+n_{-}\left(\mathbf{k}\right)}{2E_{k}}\left[\frac{2\left|\mathbf{k}\right|^{2}+m_{\nu}^{2}+\mathbf{k}\cdot\mathbf{q}}{2\mathbf{k}\cdot\mathbf{q}+\left|\mathbf{q}\right|^{2}}+\left(\mathbf{k}\to-\mathbf{k}\right)\right].
\end{eqnarray}

Furthermore, for an isotropic distribution, $n_{\pm}\left(\mathbf{k}\right)=n_{\pm}\left(\kappa\right)$ with $\kappa\equiv \left|\mathbf{k}\right|$,
one can first integrate out the angular part in Eq.~(\ref{eq:AT2}) and
obtains 
\begin{eqnarray}
{\cal A}_{{\rm bkg}}(\rho)=\frac{G_F^{2}g^1_V g^2_V}{\pi^{2}}\int_{0}^{\infty}d\kappa\frac{\kappa^{2}}{\sqrt{\kappa^{2}+m_\nu^{2}}}\left[n_{+}\left(\kappa\right)+n_{-}\left(\kappa\right)\right]\int_{-1}^{1}d\xi\frac{m_\nu^{2}+2\kappa^{2}\left(1-\xi^{2}\right)}{\rho^{2}-4\kappa^{2}\xi^{2}}\;,
\end{eqnarray}
where we have defined $\rho\equiv\left|\vec{q}\right|$ and $\xi\equiv\cos\theta$
with $\theta$ being the angel between $\mathbf{k}$ and $\mathbf{q}$.
Then the background potential is given by 
\begin{eqnarray}
V_{{\rm bkg}}(r) & = & -\int\frac{d^{3}\mathbf{q}}{\left(2\pi\right)^{3}}e^{i\mathbf{q}\cdot\mathbf{r}}{\cal A}_{\rm bkg}(\rho)=-\frac{1}{2\pi^{2}r}\int_{0}^{\infty}d\rho\rho\sin\left(\rho r\right){\cal A}_{\rm bkg}(\rho)\nonumber \\
& = & -\frac{G_F^{2}g^1_V g^2_V}{2\pi^{4}r}\int_{0}^{\infty}d\kappa\frac{\kappa^{2}}{\sqrt{\kappa^{2}+m_\nu^{2}}}\left[n_{+}\left(\kappa\right)+n_{-}\left(\kappa\right)\right]\int_{-1}^{1}d\xi\left[m_\nu^{2}+2\kappa^{2}\left(1-\xi^{2}\right)\right]\int_{0}^{\infty}d\rho\frac{\rho\sin\left(\rho r\right)}{\rho^{2}-4\kappa^{2}\xi^{2}}\nonumber \\
& = & -\frac{G_F^{2}g^1_V g^2_V}{4\pi^{3}r}\int_{0}^{\infty}d\kappa\frac{\kappa^{2}}{\sqrt{\kappa^{2}+m_\nu^{2}}}\left[n_{+}\left(\kappa\right)+n_{-}\left(\kappa\right)\right]\int_{-1}^{1}d\xi\left[m_\nu^{2}+2\kappa^{2}\left(1-\xi^{2}\right)\right]\cos\left(2\kappa r\xi\right)\nonumber \\
& = & -\frac{G_F^{2}g^1_V g^2_V}{4\pi^{3}r^{4}}\int_{0}^{\infty}\frac{d\kappa\kappa}{\sqrt{\kappa^{2}+m_\nu^{2}}}\left[n_{+}\left(\kappa\right)+n_{-}\left(\kappa\right)\right]\left[\left(1+m_\nu^{2}r^{2}\right)\sin\left(2\kappa r\right)-2\kappa r\cos\left(2\kappa r\right)\right]\;,
\end{eqnarray}
which is just Eq.~(\ref{eq:V_iso}).

\subsection{Calculation of the integral ${\cal I}(\ell,\alpha)$ in Eq.~(\ref{eq:I_def})}
Here, we calculate the integral ${\cal I}(\ell,\alpha)$ appearing in the reactor neutrino background.
Without loss of generality, we can assume
\begin{eqnarray}
	\mathbf{k}_0=E_\nu\left(0,0,1\right)\;,\quad
	\mathbf{r}=r\left(s_\alpha,0,c_\alpha\right)\;,\quad
	\mathbf{q}=\rho\left(s_\theta c_\varphi,s_\theta s_\varphi,c_\theta\right)\;,
\end{eqnarray}
where $(c_x,s_x)\equiv \left(\cos x,\sin x\right)$ have been defined. With the above coordinates, we have
\begin{eqnarray}
	\mathbf{q}\cdot \mathbf{r} =\rho r\left(s_\alpha s_\theta c_\varphi+c_\alpha c_\theta\right)\;,\quad
	\xi\equiv\frac{\mathbf{k_0}\cdot\mathbf{q}}{\left|\mathbf{k_0}\right|\left|\mathbf{q}\right|}=c_\theta\;,\quad
	\int d^3\mathbf{q}=\int_0^\infty \rho^2d\rho \int_{-1}^{1} d\xi \int_0^{2\pi} d\varphi\;. 
\end{eqnarray}
The integral in Eq.~(\ref{eq:I_def}) turns out to be
\begin{eqnarray}
	\label{eq:mid1}
	{\cal I}&\equiv&\frac{1}{E_\nu}\int d_{}^3\mathbf{q} e^{i\mathbf{q}\cdot \mathbf{r}} \frac{1-\xi^2}{\rho^2-4E_\nu^2 \xi^2}=\frac{1}{E_\nu}\int_{0}^{2\pi}d\varphi\int_{-1}^{1}d\xi\int_0^{\infty}d\rho e^{i\rho r(s_\alpha s_\theta c_\varphi+c_\alpha c_\theta)}\frac{\rho^2\left(1-\xi^2\right)}{\rho^2-4E_\nu^2\xi^2}\nonumber\\
	&=&\frac{1}{E_\nu}\left(\int_{0}^{2\pi}d\varphi\int_{0}^{1}d\xi\int_0^{\infty}d\rho+\int_{0}^{2\pi}d\varphi\int_{-1}^{0}d\xi\int_0^{\infty}d\rho\right)e^{i\rho r(s_\alpha s_\theta c_\varphi+c_\alpha c_\theta)}\frac{\rho^2\left(1-\xi^2\right)}{\rho^2-4E_\nu^2\xi^2}\;.\nonumber \\ 
\end{eqnarray}
In the second term in the bracket of Eq.~(\ref{eq:mid1}), changing the variables as $\rho\to -\rho$ and $\xi\to -\xi$ one obtains
\begin{eqnarray}
	\label{eq:mid2}
	{\cal I}&=&\frac{1}{E_\nu}\int_0^{2\pi}d\varphi\int_{0}^1 d\xi\left[\int_{0}^{\infty}d\rho e^{i\rho r(s_\alpha s_\theta c_\varphi+c_\alpha c_\theta)}+\int_{-\infty}^0 d\rho e^{i\rho r(-s_\alpha s_\theta c_\varphi+c_\alpha c_\theta)}\right]\frac{\rho^2\left(1-\xi^2\right)}{\rho^2-4E_\nu^2\xi^2}\nonumber\\
	&=& \frac{1}{E_\nu}\int_0^{\pi}d\varphi\int_{0}^1 d\xi\left[\int_{0}^{\infty}d\rho e^{i\rho r(s_\alpha s_\theta c_\varphi+c_\alpha c_\theta)}+\int_{-\infty}^0 d\rho e^{i\rho r(-s_\alpha s_\theta c_\varphi+c_\alpha c_\theta)}\right]\frac{\rho^2\left(1-\xi^2\right)}{\rho^2-4E_\nu^2\xi^2}\nonumber\\
	&+& \frac{1}{E_\nu}\int_\pi^{2\pi}d\varphi\int_{0}^1 d\xi\left[\int_{0}^{\infty}d\rho e^{i\rho r(s_\alpha s_\theta c_\varphi+c_\alpha c_\theta)}+\int_{-\infty}^0 d\rho e^{i\rho r(-s_\alpha s_\theta c_\varphi+c_\alpha c_\theta)}\right]\frac{\rho^2\left(1-\xi^2\right)}{\rho^2-4E_\nu^2\xi^2}\;.\nonumber\\
\end{eqnarray}	
In the last line of Eq.~(\ref{eq:mid2}) changing the variable $\varphi\to\varphi-\pi$ one obtains
\begin{eqnarray}
	{\cal I}&=& \frac{1}{E_\nu}\int_0^{\pi}d\varphi\int_{0}^1 d\xi\left[\int_{0}^{\infty}d\rho e^{i\rho r(s_\alpha s_\theta c_\varphi+c_\alpha c_\theta)}+\int_{-\infty}^0 d\rho e^{i\rho r(-s_\alpha s_\theta c_\varphi+c_\alpha c_\theta)}\right]\frac{\rho^2\left(1-\xi^2\right)}{\rho^2-4E_\nu^2\xi^2}\nonumber\\
	&+& \frac{1}{E_\nu}\int_0^{\pi}d\varphi\int_{0}^1 d\xi\left[\int_{0}^{\infty}d\rho e^{i\rho r(-s_\alpha s_\theta c_\varphi+c_\alpha c_\theta)}+\int_{-\infty}^0 d\rho e^{i\rho r(s_\alpha s_\theta c_\varphi+c_\alpha c_\theta)}\right]\frac{\rho^2\left(1-\xi^2\right)}{\rho^2-4E_\nu^2\xi^2}\nonumber\\
	&=&\frac{1}{E_\nu} \int_0^\pi d\varphi \int_0^1 d\xi\left(1-\xi^2\right) \int_{-\infty}^{\infty}d\rho \left[e^{i\rho r(s_\alpha s_\theta c_\varphi+c_\alpha c_\theta)}+e^{i\rho r(-s_\alpha s_\theta c_\varphi+c_\alpha c_\theta)}\right]\frac{\rho^2}{\rho^2-4E_\nu^2\xi^2}\;.\nonumber\\
\end{eqnarray}
What we have done is to change the integral to the standard form of one-dimensional Fourier transform. Then one can use the following Fourier transform:
\begin{eqnarray}
	\int_{-\infty}^\infty d\rho e^{i\rho x}\frac{\rho^2}{\rho^2-4E_\nu^2 \xi^2}=2\pi\left[\delta(x)-{\rm sgn}(x) E_\nu \xi \sin\left(2E_\nu \xi x\right)\right]\;,
\end{eqnarray}	
from which one obtains
\begin{eqnarray}
	\label{eq:generaldirect}
	{\cal I}&=&\frac{2\pi}{E_\nu} \int_0^{\pi}d\varphi\int_0^1 d\xi\left(1-\xi^2\right)\left[\delta\left(x_+\right)-{\rm sgn}\left(x_+\right)E_\nu \xi\sin\left(2E_\nu \xi x_+\right)\right.\nonumber\\
	&&\left.+\delta\left(x_-\right)-{\rm sgn}\left(x_-\right)E_\nu \xi\sin\left(2E_\nu \xi x_-\right)\right]\nonumber\\
	&=&\frac{2\pi}{E_\nu}\int_{-1}^1 d\xi\left(1-\xi^2\right)\int_0^{\pi}d\varphi\left[\delta\left(x_+\right)-{\rm sgn}\left(x_+\right)E_\nu \xi\sin\left(2E_\nu \xi x_+\right)\right]
\end{eqnarray}
where $x_{\pm}\equiv r(\pm s_\alpha s_\theta c_\varphi+c_\alpha c_\theta)$. The first term in Eq.~(\ref{eq:generaldirect}) involving $\delta$ function can be analytically integrated out
\begin{eqnarray}
	&&\frac{2\pi}{E_\nu}\int_{-1}^{1}d\xi\left(1-\xi^2\right)\int_0^{\pi}d\varphi \delta\left[r\left(s_\alpha s_\theta c_\varphi+c_\alpha c_\theta\right)\right]\nonumber\\
	&=&\frac{2\pi}{rE_\nu}\int_{-1}^1d\xi\left(1-\xi^2\right)\int_0^\pi d\varphi\delta\left(s_\alpha\sqrt{1-\xi^2}c_\varphi+c_\alpha \xi\right)\nonumber\\
	&\overset{t=c_\varphi}{=}&\frac{2\pi}{\ell s_\alpha}\int_{-1}^1 d\xi\left(1-\xi^2\right)\frac{1}{\sqrt{1-\xi^2}}\int_{-1}^{1}\frac{dt}{\sqrt{1-t^2}}\delta\left(t+\frac{\xi}{\sqrt{1-\xi^2}}\cot\alpha\right)\nonumber\\
	&=&\frac{2\pi}{\ell s_\alpha}\int_{-\frac{1}{\sqrt{1+\cot^2\alpha}}}^{\frac{1}{\sqrt{1+\cot^2\alpha}}} d\xi \frac{1-\xi^2}{\sqrt{1-\left(1+\cot^2\alpha\right)\xi^2}}\nonumber\\
	&=& \frac{\pi^2}{2 \ell}\left(3+\cos2\alpha\right)\;,
\end{eqnarray}
where we have defined the dimensionless quantity $\ell\equiv r E_\nu$.
The second term in Eq.~(\ref{eq:generaldirect}) cannot be analytically integrated. So finally one obtains the directional integral
\begin{eqnarray}
	{\cal I}\left(\ell,\alpha\right)&=&\frac{\pi^2}{2\ell}\left(3+\cos2\alpha\right)\nonumber\\
	&&-2\pi  \int_{-1}^1 d\xi\, \xi\left(1-\xi^2\right)\int_0^\pi d\varphi\,{\rm sgn}\left(s_\alpha \sqrt{1-\xi^2} c_\varphi+c_\alpha \xi\right)\sin\left[2\ell \xi\left(s_\alpha \sqrt{1-\xi^2} c_\varphi+c_\alpha \xi\right)\right]\nonumber\\
	&=&\frac{\pi^2}{2\ell}\left(3+\cos2\alpha\right)-2\pi\int_{-1}^1d\xi\,\xi \left(1-\xi^2\right)\int_0^\pi d\varphi\, \sin\left(2\ell\xi\left|s_\alpha \sqrt{1-\xi^2} c_\varphi+c_\alpha \xi\right|\right)\;,
\end{eqnarray}
which is the result of Eq.~(\ref{eq:I_full}). Then the directional background potential is given by
\begin{eqnarray}
	V_{\rm bkg}(r,\alpha)=-\frac{g^1_V g^2_V}{\pi^3}G_F^2 \Phi_0 E_\nu^2 \times {\cal I}\left(rE_\nu,\alpha\right)\;.
\end{eqnarray}

\section{Energy distribution function with a finite spread}
\label{app:spread}
In this appendix, we consider a directional neutrino flux with a finite energy spread instead of the monochromatic case we considered in the main text in Eq.~(\ref{eq:x-1}). 

Neglecting the neutrino mass, which is much smaller than the typical energy of neutrino flux,
the general directional neutrino flux can be written as
\begin{eqnarray}
	\label{eq:spread}
n_{\pm}\left(\mathbf{k}\right)=\left(2\pi\right)^{3}f(E)\delta\left(\hat{\mathbf{k}}-\hat{\mathbf{k}}_{0}\right)\;,
\end{eqnarray}
where $\hat{\mathbf{k}}$ denotes the unit vector of the three momentum $\mathbf{k}$, while $\hat{\mathbf{k}}_0$ represents a certain direction. Without loss of generality, we take $\hat{\mathbf{k}}_0=(0,0,1)$ and such that the delta function enforces $\mathbf{k}=(0,0,E)$.
The energy distribution function $f(E)$ should satisfy the normalization condition:
\begin{eqnarray}
	\label{eq:renormalization}
\int \frac{d^3 \mathbf{k}}{\left(2\pi\right)^3} n_{\pm}\left(\mathbf{k}\right)=\Phi_0\;,
\end{eqnarray}
with $\Phi_0$ being the total flux of neutrinos. 

For example, a Gaussian-like distribution reads
\begin{eqnarray}
	\label{eq:Gaussian}
f_g(E)=\frac{\Phi_0}{2\pi B} {\rm exp}\left[-\frac{\left(E-E_0^2\right)}{2\sigma_E^2}\right]\;,
\end{eqnarray}
where $E_0^{}$ is the mean energy and $\sigma_E$ denotes the spread of energy. The normalization factor $B$ is given by
\begin{eqnarray}
B&=& \int_{0}^{\infty} dE E^2 {\rm exp}\left[-\frac{\left(E-E_0\right)^2}{2\sigma_E^2}\right]\nonumber\\
&=&E_0 \sigma_E^2\,{\rm exp}\left(-\frac{E_0^2}{2\sigma_E^2}\right)+\sqrt{\frac{\pi}{2}}\sigma_E\left(E_0^2+\sigma_E^2\right)\left[1+{\rm Erf}\left(\frac{E_0}{\sqrt{2}\sigma_E}\right)\right]\nonumber\\
&=& \sqrt{2\pi}E_0^2\sigma_E+{\cal O}\left(\sigma_E^3\right)\;.
\end{eqnarray}
It can be verified explicitly that the distribution in Eq.~(\ref{eq:Gaussian}) satisfies the normalization in Eq.~(\ref{eq:renormalization}). In particular, in the limit of $\sigma_E\to 0$ one obtains 
\begin{equation}
f_g(E)\delta\left(\hat{\mathbf{k}}-\hat{\mathbf{k}}_{0}\right)\to \frac{\Phi_0}{2\pi E_0^2}\delta\left(E-E_0\right)\delta\left(\hat{\mathbf{k}}-\hat{\mathbf{k}}_{0}\right)=\delta^3\left(\mathbf{k}-\mathbf{k}_0\right)\Phi_{0}\;,
\end{equation}
which reduces to the monochromatic case in Eq.~(\ref{eq:x-1}).

Below we compute the background potential without the specific form of $f(E)$ for the purpose of generality. Substituting Eq.~(\ref{eq:spread}) in Eq.~(\ref{eq:Abkg}), one obtains
\begin{eqnarray}
{\rm \cal A}_{\rm bkg}(\mathbf{q})=2G_F^2 g_V^1 g_V^2\int d^3 \mathbf{k}\frac{f(E)}{E}\delta\left(\hat{\mathbf{k}}-\hat{\mathbf{k}}_{0}\right)\left[\frac{2\left|\mathbf{k}\right|^{2}+\mathbf{k}\cdot\mathbf{q}}{2\mathbf{k}\cdot\mathbf{q}+\left|\mathbf{q}\right|^{2}}+\left(\mathbf{k}\to-\mathbf{k}\right)\right]\;.
\end{eqnarray}
Then using the decomposition
\begin{equation}
\int d^3 {\mathbf{k}}\delta\left(\hat{\mathbf{k}}-\hat{\mathbf{k}}_{0}\right) f(E)=2\pi \int_{-1}^1 dz \delta\left(z-1\right)\int_0^{\infty} dE E^2 f(E)\;,
\end{equation}
where $z\equiv \hat{\mathbf{k}}\cdot\hat{\mathbf{k}}_0$, we have
\begin{eqnarray}
{\rm \cal A}_{\rm bkg}(\mathbf{q})&=&4\pi G_F^2 g_V^1 g_V^2\int_{-1}^1 dz \delta\left(z-1\right)\int_{0}^{\infty}dE E f(E)\left[\frac{2E^2+E\rho \xi}{2E\rho\xi+\rho^2}+\frac{2E^2-E\rho \xi}{-2E\rho\xi+\rho^2}\right]\nonumber\\
&=& 16 \pi G_F^2 g_V^1 g_V^2 \int_{0}^{\infty}dE E^3 f(E)\frac{1-\xi^2}{\rho^2-4E^2\xi^2}\;.
\end{eqnarray}
Notice that $\rho\equiv \left|\mathbf{q}\right|$ and $	\xi\equiv\frac{\mathbf{k}\cdot\mathbf{q}}{\left|\mathbf{k}\right|\left|\mathbf{q}\right|}$ have been defined. The background potential turns out to be
\begin{eqnarray}
	\label{eq:vbkgspread}
V_{{\rm bkg}}(\mathbf{r})&=&-\int\frac{d^{3}\mathbf{q}}{\left(2\pi\right)^{3}}e^{i\mathbf{q}\cdot\mathbf{r}}{\cal A}_{{\rm bkg}}(\mathbf{q})= -\frac{2}{\pi^2}G_F^2g_V^1g_V^2\int_{0}^{\infty}dE E^3 f(E)\int d^{3}\mathbf{q}e^{i\mathbf{q}\cdot\mathbf{r}}\frac{1-\xi^2}{\rho^2-4E^2\xi^2}\nonumber\\
&=&-\frac{2}{\pi^2}G_F^2 g_V^1 g_V^2 \int_{0}^{\infty}dE E^4 f(E) \,{\cal I}\left(Er,\alpha\right)\;,
\end{eqnarray}
where the dimensionless integral is defined as 
\begin{eqnarray}
{\cal I}\left(Er,\alpha\right)\equiv \frac{1}{E}\int d^{3}\mathbf{q}e^{i\mathbf{q}\cdot\mathbf{r}}\frac{1-\xi^2}{\rho^2-4E^2\xi^2}\;,
\end{eqnarray}
whose result has been given by Eq.~(\ref{eq:I_full}) with the substitution $\ell=Er$. In particular, in the monochromatic limit, the background potential reduces to Eq.~(\ref{eq:anisotropicpotential}):
\begin{eqnarray}
\qquad f(E)\to \frac{\Phi_0}{2\pi E_0^2}\delta\left(E-E_0\right)\;,\qquad
V_{\rm bkg}(\mathbf{r})\to-\frac{1}{\pi^3}G_F^2g_V^1g_V^2\Phi_0E_0^2\,{\cal I}\left(E_0 r,\alpha\right)\;.
\end{eqnarray}

To sum up, the background potential in a directional neutrino flux with an arbitrary finite energy spread is given by Eq.~(\ref{eq:vbkgspread}), with the integral ${\cal I}$ being computed in Eq.~(\ref{eq:I_full}).

\section{Flavor- and material-dependence of the background potential}
\label{app:ZA}

In Sec.~\ref{sec:exp} we have neglected the effects of neutrino flavors and materials of test bodies when computing the directional neutrino background potential. Here we compute a complete expression for the neutrino force between two objects with masses $m_1$ and $m_2$, as a function of the background neutrino flavor distribution and their respective atomic and mass numbers. We present the expression under the following assumptions:
\begin{enumerate}
    \item Let the masses be pure elements of atomic numbers $Z_1$ and $Z_2$ respectively. Let their mass numbers be $A_1$ and $A_2$ respectively. 
    \item We further assume that the mass of the objects are constituted entirely by the masses of the neutrons and protons in the object, i.e, we ignore electron mass $m_e \ll m_p \approx m_n$, where the subscripts $p$ and $n$ stand for proton and neutron respectively.
    \item We assume the massless limit for the neutrinos, where the mass eigenstates are identical to the flavor eigenstates.
    \item We see in the text how finite spread of the masses weakens the $1/r$ behavior of the neutrino background potential. In this appendix, we assume that the angular spread $\alpha^2 \ll 1/\Delta(E_{\nu} r)$, where $r$ is the distance between the masses and $E_{\nu}$ is defined in the text [see Eq.~(\ref{eq:alpha-crit})].
\end{enumerate}

Given masses $m_i$ (for $i=1,2$), the number of protons, neutrons and electrons in each mass is given by 

\beq N_p^i  \equiv {m_i Z_i \over A_i m_p} = N_e ^i \;, \qquad N_n^i \equiv {m_i (A_i - Z_i) \over A_i m_p}\;.  \eeq

The effective $g_V$'s for each mass can then be computed simply by adding up the $g_V$'s of each of the constituent species and multiplying by the corresponding number of that species in the mass. The effective $g_V$ depends on which neutrino is being exchanged between the masses. For instance, when the neutrinos exchanged are electron neutrinos, we get:

\beq g_{Ve}^i = N_p^i (1/2 - 2 s_W^2) + N_e^i (1/2+ 2 s_W^2) - N_n^i/2 = N_p^i - N_n^i/2 \;,\eeq
where $s_W$ is the sine of the Weinberg angle $\theta_W$.
For other neutrinos being exchanged the effective coupling is:
\beq g_{V\mu/\tau}^i = N_p^i (1/2 - 2 s_W^2) + N_e^i (-1/2+ 2 s_W^2) - N_n^i/2 =  - N_n^i/2 \;.\eeq
Note that, in the presence of electron neutrino background, the electrons in the material need to be considered when calculating the force.

In the end, the neutrino background potential between the two masses is given by (we have taken $\alpha=0$ like what we did in Eq.~(\ref{eq:V_for_plot})] in accordance with assumption 4 above):
\beq V_{\text{bkg}} (r) = -{G_F^2 \Phi E_{\nu} \over \pi r } \left[n_e g_{Ve}^1 g_{Ve}^2 + \left(1-n_e\right) g_{V\mu/\tau}^1 g_{V\mu/\tau}^2 \right],\eeq
where $n_e$ is the fraction of electron neutrinos in the flux $\Phi$. 
After some algebra this can be written as:
\beq V_{\text{bkg}} (r) = -{G_F^2 \Phi E_{\nu} \over \pi r } {m_1 m_2 \over m_p^2} \times f(A_1,A_2,Z_1,Z_2,n_e)\;, \label{eq:vbkgf}\eeq
where
\beq f(A_1,A_2,Z_1,Z_2,n_e) = {1 \over 4} \left[ n_e \left( {3 Z_1 \over A_1} - 1 \right)\left( {3Z_2 \over A_2} - 1 \right)+ (1-n_e) \left(1-{Z_1 \over A_1} \right) \left(1- {Z_2 \over A_2
} \right)\right].\eeq

The net potential between these two masses is therefore given by:
\beq V_{\rm net} = V_{\rm grav}  + V_{\rm bkg} = -{m_1 m_2 \over r} \left[ G_N + {G_F^2 \Phi E_{\nu} \over \pi m_p^2} f(A_1,A_2,Z_1,Z_2,n_e) \right], \eeq
where $G_N$ is the gravitational constant.
We have ignored the  $1/r^5$ term from the vacuum neutrino force since at the distances we are talking about that force is negligible. Note that the Weinberg angle does not feature in our final expression for the neutrino force. 

The ratio of the neutrino force to the gravitational force between these two masses at some distance $r\gg E_\nu^{-1}$ is independent of $r$,

\beq {V_{\rm bkg}(r)\over V_{\rm grav}(r)} = {G_F^2 \Phi E_{\nu}f(A_1,A_2,Z_1,Z_2,n_e) \over \pi G_N m_p^2 }\;.\eeq

Below we mention some special cases:

\begin{enumerate}
    \item Consider the special case $Z_1 = Z_2=Z, A_1 = A_2=A$ and $n_e = 1$, i.e, the background is purely electron neutrino. In this case,  the ratio reads:

    \beq {V_{\rm bkg}(r)\over V_{\rm grav}(r)} = {G_F^2 \Phi E_{\nu}  \over 4 \pi G_N m_p^2 } \left( {3 Z \over A} - 1 \right)^2. \eeq
    Note that this ratio is maximized when $Z=A$, i.e, for Hydrogen.
    
    Putting in the numbers we get (using $\Phi \sim 10^{14} {\rm cm}^{-2}{\rm s}^{-1}$ and $E_{\nu} \sim 1$ MeV):
    \beq {V_{\rm bkg}(r)\over V_{\rm grav}(r)} \sim 10^{-13}\;. \eeq 
    The gravitational force is thus 13 orders of magnitude greater than the neutrino background force in this limit. This corresponds to the purple line (reactor 10m) in Fig.~\ref{fig:Exp}.
    
    \item Consider the special case $Z_1 = Z_2=Z, A_1 = A_2=A$ and $n_e = 0$, i.e, the background is purely muon/tau neutrino. In this case we note that force is entirely due to the number of neutrons in the masses, and the ratio:

    \beq {V_{\rm bkg}(r)\over V_{\rm grav}(r)} = {G_F^2 \Phi E_{\nu}  \over 4 \pi G_N m_p^2 } \left( 1-{Z \over A} \right)^2.\eeq
    In the special case of Hydrogen we see that we shall not find any additional force due to background neutrinos. However in other elements we can see this effect. 
\end{enumerate}
To finish this section, we show how the force varies for different materials. For a given $\Phi$ and $E_{\nu}$, and assuming that $A \approx 2Z$ as is usually the case for most elements, we have,
 \beq {\delta V_{\rm bkg} \over V_{\rm bkg}}  \approx 4 \left(4n_e-1\right) \delta_{Z/A}\;, \label{eq:dVoverV} \eeq
where $\delta_{Z/A}$ refers to the variation of $Z/A$ for different materials, as in Eq.~\eqref{eq:x}.

\end{appendix}

\bibliographystyle{JHEP}
\bibliography{ref}
\end{document}